\documentclass[aos,preprint]{imsart}
\usepackage[authoryear,round]{natbib}
\usepackage{amsmath,amssymb,amsthm,booktabs,microtype,graphicx,fix-cm}
\usepackage{xcolor}
\usepackage{float}
\floatstyle{ruled}
\newfloat{algorithm}{tbp}{loa}
\floatname{algorithm}{Algorithm}
\usepackage{pdfpages}
\usepackage[hidelinks]{hyperref}
\DeclareFontShape{T1}{ptm}{m}{scit}{<->ssub * ptm/m/sc}{}
\newtheorem{theorem}{Theorem}
\newtheorem{lemma}[theorem]{Lemma}
\newtheorem{corollary}[theorem]{Corollary}
\newtheorem{proposition}[theorem]{Proposition}
\newcommand{\E}{\mathcal E}
\newcommand{\V}{\mathcal V}
\newcommand{\K}{\mathcal K}
\newcommand{\argmin}{\operatorname*{arg\,min}}
\newcommand{\supp}{\operatorname{supp}}
\newcommand{\dist}{\operatorname{dist}}

\theoremstyle{plain}
\newtheorem{assumption}{Assumption}

\newtheorem*{problem}{Problem}
\theoremstyle{definition}
\newtheorem{definition}{Definition}
\newtheorem{example}{Example}
\newtheorem{remark}{Remark}
\begin{document}
\begin{frontmatter}
\title{Identifying and estimating sparsest representations of linear Gaussian causal models with hidden confounding}
\runtitle{Sparsest representations of Gaussian causal models}
\begin{aug}
\author[A]{\fnms{Amir}~\snm{Asiaee}\ead[label=e1]{amir.asiaeetaheri@vumc.org}}
\author[A]{\fnms{Samhita}~\snm{Pal}\ead[label=e2]{samhita.pal@vumc.org}}
\author[B]{\fnms{James}~\snm{O'Quinn}\ead[label=e3]{joquinn4@jh.edu}}
\author[C]{\fnms{James P.}~\snm{Long}\ead[label=e4]{jplong@mdanderson.org}}
\address[A]{Department of Biostatistics, Vanderbilt University Medical Center\printead[presep={,\ }]{e1,e2}}
\address[B]{Department of Computer Science, Johns Hopkins University\printead[presep={,\ }]{e3}}
\address[C]{Department of Biostatistics, The University of Texas MD Anderson Cancer Center\printead[presep={,\ }]{e4}}
\end{aug}
\runauthor{A. Asiaee et al.}
\begin{abstract}
We study identification and estimation of linear Gaussian structural
equation models with sparse hidden confounding when both the directed
structure and the error covariance are unknown. Although observationally
equivalent representations can form continua, minimizing the joint count
of directed and bidirected edges generically yields a finite set of
regular representations containing the generating parameter whenever the
generating graph is bow-free, or more generally has a generically
finite-to-one covariance map. Every such representation is
algebraically, hence Markov, equivalent to the generating graph, even
when competitors contain bows. On a compact domain, a broad family of
continuous saturated penalties minimizes the population likelihood
exactly at these sparsest representations, whereas an $\ell_1$ penalty
can leave the generating Markov class, even for standardized variables.
For fixed dimension, thresholded approximate global minimizers
consistently select the generating algebraic class, and refitting
inherits the convergence rate of the covariance estimator. We establish
explicit separation conditions for class recovery and an
information-theoretic lower bound. Finally, a proximal alternating
algorithm with exact vertex-order search converges globally to critical
points and achieves one-cycle support identification and geometric
convergence locally, with a computable a posteriori certificate;
exhaustive three- and four-variable experiments illustrate selection and
local-optimization traps.
\end{abstract}
\end{frontmatter}
\section{Introduction}
\label{sec:intro}
A common way to handle hidden confounding in causal structure
learning is to treat it as pervasive. A few latent factors affect many
observed variables, and their low-rank contribution is estimated and
removed before or during structure learning
\citep{cpw2012,frot,rsvp,decam}. In many applications, however, each
hidden cause affects only a few variables, and its effect is better
described by a few correlations between structural errors. This is the
setting of the present paper.

Hidden confounding of this kind makes causal structure ambiguous. An
association between two observed variables may arise from a causal effect of one
variable on the other, from an unobserved common cause, or from both, and in a linear
Gaussian model these explanations can induce exactly the same covariance
matrix, even with unlimited data. We study what remains identifiable when
neither the directed structure nor the pattern of confounding is known.
Our answer is to target the sparsest representations of the observed
covariance, counting both directed effects and correlations between
structural errors. A unique causal graph is generally not identifiable,
but we show that this rule recovers the algebraic equivalence class of
the generating graph under generic conditions.

We consider the centered linear Gaussian structural equation model on $p$ variables,
\begin{equation}\label{eq:model}
 X=B^T X+\epsilon,\qquad \epsilon\sim N(0,\Omega),\qquad
 \Phi(B,\Omega)=(I-B)^{-T}\Omega(I-B)^{-1}.
\end{equation}
Here $B$ is a coefficient matrix with acyclic directed
support, and $\Omega$ is a positive definite error covariance that may have
nonzero off-diagonal entries. A nonzero coefficient $B_{ij}$ represents a
directed edge $i\to j$, and a nonzero off-diagonal entry $\Omega_{ij}$
represents a bidirected edge $i\leftrightarrow j$, so a parameter
$\theta=(B,\Omega)$ defines an acyclic directed mixed graph on the $p$
variables. Its directed edges are the direct causal effects
among the observed variables, and its bidirected edges mark the pairs
of variables that share a hidden cause, so learning the causal structure
means recovering both kinds of edge. We call the pair $\theta^*$ that generated the data the
generating parameter, its graph $G^*$ the generating graph, and
$\Sigma^*=\Phi(\theta^*)$ the population covariance. A representation of
$\Sigma^*$ is any $\theta$ with $\Phi(\theta)=\Sigma^*$. Every
covariance has the representation $B=0$, $\Omega=\Sigma^*$, which
attributes all dependence to hidden causes, so the set of all
representations says little about the causal structure, and we must
decide which representations to recover.

\begin{figure}[t]
\centering
\includegraphics[width=\linewidth]{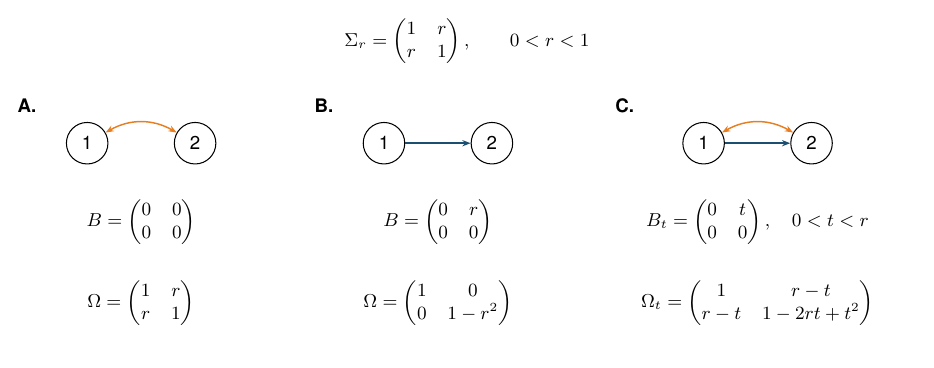}
\caption{\textbf{Different Graphs Representing the Same Covariance.}
For the standardized covariance $\Sigma_r$ in Example~\ref{ex:two_variable},
\textbf{A.} a single bidirected edge and \textbf{B.} a single directed edge
both give exact one-edge representations. \textbf{C.} For $0<t<r$,
the bowed graph also represents $\Sigma_r$ exactly, with two edges.
The matrices beneath each graph show how it allocates dependence between $B$ and $\Omega$.}
\label{fig:admg}
\end{figure}

The ambiguity already appears for two variables
(Example~\ref{ex:two_variable} and Figure~\ref{fig:admg}). Section~\ref{sec:l1} returns to this example to
compare edge counts with the sizes of parameters, and in
Remark~\ref{rem:weights} we explain why directed and bidirected edges are
counted equally.

\begin{example}[Two correlated variables]\label{ex:two_variable}
Let $\Sigma_r=\left(\begin{smallmatrix}1&r\\r&1\end{smallmatrix}\right)$
with $0<r<1$, and allow any acyclic mixed graph on two vertices. The family
\[
B_t=\begin{pmatrix}0&t\\0&0\end{pmatrix},\qquad
\Omega_t=\begin{pmatrix}1&r-t\\r-t&1-2rt+t^2\end{pmatrix}
\]
represents $\Sigma_r$ exactly for every real $t$, with $\Omega_t$
positive definite because $\det\Omega_t=1-r^2>0$. At $t=0$ a single
bidirected edge carries the dependence, at $t=r$ a single directed edge
$1\to2$ does, and every other $t$ uses both edges on the same pair,
which we call a \emph{bow}. Exchanging the roles of the two
variables gives a second family of the same form, with directed edge
$2\to1$; its member with $B_{21}=r$, $\Omega_{11}=1-r^2$, and $\Omega_{22}=1$ is a third one-edge representation. The representations of $\Sigma_r$
therefore have five graphs, the three one-edge graphs and the two bows.
The three one-edge representations are the sparsest, and the bowed ones,
with two edges each, are not. The set of all exact representations, sparse or not, is the \emph{covariance fiber} and contains both families. Each family is a curve of representations, or a \emph{continuum}, because it has infinitely many members that change continuously with $t$.
\end{example}

Example~\ref{ex:two_variable} rules out recovery of a unique
graph, but its three sparsest graphs impose the same constraints on the
covariance. This suggests a weaker identification question. Does every
sparsest representation of a generic generating covariance belong to the
algebraic equivalence class of the generating graph? Two graphs are
algebraically equivalent when their covariance models satisfy the same
polynomial equality constraints (Section~\ref{sec:equivalence}).
Algebraic equivalence implies Markov equivalence, and hence agreement on
conditional independences, and it can be strictly finer, as the
four-variable example of Section~\ref{sec:verma} shows.

Our first result answers this question affirmatively when the generating
graph is bow-free, meaning that no pair of variables is joined by both a
directed and a bidirected edge, and its parameter is generic
(Theorem~\ref{thm:ident}). More generally, it holds whenever the
covariance map of the generating graph is generically finite-to-one. We
do not require the competing graphs to be bow-free. Every exact
representation with as few edges as the generating graph belongs to its
algebraic equivalence class, even if that representation contains bows.
The reason is a dimension count. A bow-free graph with $e$ edges produces
a covariance model of dimension $p+e$, while any mixed graph with $e$
edges produces one of dimension at most $p+e$, so a competing graph that
reproduces a generic covariance of the generating model with no more
edges must have the same covariance variety. Although the set of all
exact representations, the covariance fiber, can be a continuum, its
subset of sparsest representations is generically finite
(Corollary~\ref{cor:regular}).

Identification alone does not give an estimator. Our practical starting
point is the DECOR method of \citet{decor}, which estimates the
coefficient matrix and the error covariance jointly by minimizing a
Gaussian likelihood with an $\ell_1$ penalty on $B$ and on the
off-diagonal part of $\Omega$, under the smooth acyclicity constraint of
NOTEARS \citep{notears2018}. That work reported promising results in
simulation studies but gave no statistical or computational guarantees
for its estimator. We keep its joint formulation and replace the $\ell_1$
penalty by a saturated penalty, which assigns the same cost to every edge
coordinate above a threshold and therefore counts edges instead of adding
up their magnitudes. We prove that, on a compact parameter domain, the minimizers of the penalized population likelihood are exactly the sparsest representations, even
when the covariance fiber is infinite. From data, accurate approximate
global minimizers recover a minimum-edge support after thresholding, and
refitting attains the parametric rate. An $\ell_1$ penalty, by contrast,
need not favor the sparsest representation, and even with standardized
variables it can select a graph outside the Markov equivalence class of
the generating graph (Section~\ref{sec:l1}).

These statistical guarantees concern global minimizers of a nonconvex
objective, and they do not follow from the convergence of a local
algorithm. We therefore treat computation separately
(Figure~\ref{fig:guarantees}). We develop a proximal alternating method
that updates $B$ and $\Omega$ in turn and computes each update of $B$
exactly by searching over vertex orders. From any feasible start the
method converges to a critical point, and near a regular sparsest
representation, one at which the covariance map of its graph has
a derivative of full rank, it identifies the support and converges geometrically. A
computable a posteriori certificate verifies an attracting neighborhood
around a computed output, but it certifies neither global optimality nor
membership in the generating class. Our three- and four-variable experiments show why this distinction matters, since they exhibit bowed local minima and a strong
dependence on the starting point.

\subsection{Related work}\label{sec:related}
Path diagrams with directed and bidirected edges go back to the
path coefficients of Sewall Wright \citep{wright1934}. When the graph is
known, a well-developed theory decides whether the covariance determines
$B$ and $\Omega$ \citep{britopearl2002,fdd2012,drtonweihs2016,dfs2011}.
When the graph is unknown, additional structure can identify it, such as
non-Gaussian errors \citep{lingam,wangdrton2023}, equal or known error
variances \citep{equalvar,loh2014} or nonlinear mechanisms
\citep{anm,defuse}. In the linear Gaussian model with unrestricted error
correlations that we study, none of this structure is available, and the
covariance can determine the graph only up to an equivalence class.
Markov equivalence compares the conditional independences that two
graphs imply \citep{ars2009,zhang2008,rfci2012}, algebraic equivalence
compares all polynomial equality constraints on their covariance models
and is finer \citep{ommen2017,ommen2024}, and distributional equivalence
compares the covariance models themselves \citep{nowzohour}.

Preferring
the sparsest representation is a familiar way to choose within such a
class, both for directed acyclic graphs
\citep{vandegeer,raskutti,solus2021} and with latent variables
\citep{bernstein2020,dcd,dong2026}. Penalties that are constant
beyond a threshold \citep{scad2001,mcp2010,tlp2012} turn this preference
into an estimator \citep{aragamzhou2015,yuan2019,deng,jin2026}. In the partially observed linear Gaussian setting, \citet{dong2026} use model dimension to characterize an algebraic equivalence class under a generalized faithfulness condition. Our identification argument also compares
dimensions, but it analyzes the joint edge count of a generic generating
graph and allows competing graphs with bows.

Mixed
graphs have been learned by greedy, exact and continuous search
\citep{nowzohour,rantanen2021,dcd,ashman2023,spot}, and the
pervasive models cited at the start of this section remove a low-rank
component instead of modeling sparse error correlations. What has been missing is a statistical
theory that says which class the sparsest representations recover when
the errors are correlated and the graph, including its bows, is unknown,
and how a penalized likelihood estimates that class. We compare our
results with the closest of these works where each result appears
(Remarks~\ref{rem:sparsest} and~\ref{rem:finite_fiber},
Sections~\ref{sec:l1}, \ref{sec:computation} and~\ref{sec:discussion}).
\subsection{Our contributions}
\label{sec:contributions}
Given a covariance input $S$, we minimize
\[
 \ell_S(\theta)+\lambda P_\delta(\theta),\qquad \theta=(B,\Omega),
\]
over acyclic coefficient matrices and error covariances in a compact
domain, where $\ell_S$ is the Gaussian negative log likelihood and
the saturated penalty $P_\delta$ assigns penalty value one to each edge coordinate of magnitude at least $\delta$ and a smaller value to smaller ones
(Section~\ref{sec:estimation}). We search for minimizers by alternating between updates of the
coefficient matrix and of the error covariance
(Section~\ref{sec:computation}), and we then set small entries to zero
and maximize the likelihood again on the edges that remain.

As in learning directed acyclic graphs, the covariance is unknown, the edge
count is discontinuous and the set of representations is nonconvex.
Correlated errors add four difficulties of their own. First, the population target is a set of representations whose graphs can mix directed and bidirected edges and may contain bows. Second, the covariance fiber can
be a continuum, even though the target set inside it is generically
finite. Third, setting a small entry of $\Omega$ to zero can destroy its
positive definiteness, whereas setting an entry of $B$ to zero keeps $B$
acyclic. Fourth, a nondiagonal error covariance couples the regressions
of all variables in the likelihood, so exact coefficient fits do not
separate across variables, and the algorithm must also choose an
ordering of the variables.

Our main results are the following.
First, for a generic parameter of a bow-free generating graph, the
sparsest representations form a finite set that contains the generating
parameter, and all of their graphs are algebraically, hence Markov,
equivalent to the generating graph, although these graphs and their
competitors may contain bows (Section~\ref{sec:identification}).

Second, on a compact parameter domain and for small tuning parameters, the penalized population likelihood is minimized exactly at the sparsest representations, without requiring the covariance fiber to be finite as earlier results do. After thresholding, accurate approximate global minimizers of the sample objective select, with probability tending to one, a graph in the generating algebraic class; refitting on that graph attains the parametric rate for the sample covariance and inherits the convergence rate of other consistent covariance estimates (Section~\ref{sec:estimation}).

Third, we show why saturation matters. An $\ell_1$ penalty can favor
representations with redundant edges and can select a graph that is not
even Markov equivalent to the generating graph, even with standardized variables. For the saturated penalty we give an explicit separation
condition for class selection and show that no estimator selects the
class when the correlations are too weak, so that the critical
correlation is of order $n^{-1/2}$ up to a slowly growing factor
(Section~\ref{sec:separation}).

Fourth, the alternating algorithm converges from any feasible start to
a critical point. Started near a regular sparsest representation, it finds the correct graph after one cycle and converges geometrically, and a computable certificate verifies local convergence around a given output (Section~\ref{sec:computation}).

Exhaustive computations on three and four variables agree with the
selection results, while random-start experiments expose the local
minima that separate the statistical guarantees from the computational
ones (Section~\ref{sec:numerics}). Our statistical results are for a
fixed number of variables, and the alternating algorithm is guaranteed
to reach a sparsest representation only from a nearby start.

\subsection{Organization}
\label{sec:organization}
Section~\ref{sec:model} defines the covariance fiber, the
minimum-edge target and the equivalence relations between graphs.
Section~\ref{sec:identification} proves generic identification,
Sections~\ref{sec:estimation} and~\ref{sec:separation} develop the
statistical guarantees and compare penalties, Section~\ref{sec:computation}
analyzes the alternating algorithm and its certificate,
Section~\ref{sec:numerics} reports numerical experiments, and
Section~\ref{sec:discussion} discusses limitations and extensions. The Supplementary Material \citep{supplement} contains the proofs of
all results, numbered as in this article. Its Sections S1--S4 cover
identification, Sections S5--S13 estimation, separation and the $\ell_1$
penalty, Sections S14--S17 computation and the certificate, and
Section S18 the numerical details.

\section{Background and problem formulation}\label{sec:model}
In this section we fix the graph terminology, define the
estimation target, and separate the covariance model of a graph from the
algebraic set that we use in the identification argument.

\subsection{Linear structural equation models with correlated errors}
In model~\eqref{eq:model}, a nonzero coefficient $B_{ij}$ corresponds to
a directed edge $i\to j$, and a nonzero off-diagonal entry $\Omega_{ij}$
corresponds to a bidirected edge $i\leftrightarrow j$. An
\emph{acyclic directed mixed graph} (ADMG) \citep{richardson2003} on the
vertices $1,\dots,p$ is a set of directed edges $i\to j$ and bidirected
edges $i\leftrightarrow j$ that contains no directed cycle. A graph
records only which entries of $B$ and $\Omega$ may be nonzero, not their
values; a representation $\theta=(B,\Omega)$ supplies the values. We require $B_{ii}=0$ and
$\Omega\succ0$, and call the graph bow-free when no pair carries both
a directed and a bidirected edge, a condition introduced by
\citet{britopearl2002}, as illustrated in Figure~\ref{fig:admg}.

The support graph $G(\theta)$ of a representation
$\theta=(B,\Omega)$, also written $\supp(\theta)$, contains precisely the edges corresponding to nonzero
entries of $B$ and $\Omega$. For a graph $H$, $e(H)$ counts its directed
and bidirected edges, and the cost of the representation $\theta$ is the
joint edge count of its support graph, $c(\theta)=e(G(\theta))$, that is,
\begin{equation}\label{eq:cost}
 c(\theta)=\sum_{i\ne j}\mathbf1\{B_{ij}\ne0\}
            +\sum_{i<j}\mathbf1\{\Omega_{ij}\ne0\}.
\end{equation}
We count each unordered off-diagonal error covariance entry
once and leave the $p$ error variances $\Omega_{ii}$ free and unpenalized,
so the cost measures how sparse a representation is without specifying
the number of hidden causes or how they load on the variables. The cost
does not change when the variables are rescaled. The penalties on edge magnitudes and the parameter bounds introduced later do change under rescaling, so we
state them for a fixed scale of the variables.

For a matrix $M$ we write $\|M\|_F$ for the Frobenius norm and $\|M\|$
for the spectral norm, and for a representation we set
$\|\theta\|=\|\theta\|_F=(\|B\|_F^2+\|\Omega\|_F^2)^{1/2}$. We write
$\operatorname{pr}$ for probability, because the letter $P$ denotes the
penalty of Section~\ref{sec:estimation}.

\subsection{The covariance fiber and the estimation target}\label{sec:fiber}

\begin{definition}[Covariance fiber and minimum-edge fiber]\label{def:fiber}
Let $\mathcal A_0$ be the unrestricted class of all pairs $(B,\Omega)$
with $B$ acyclic and $\Omega$ positive definite, and call any subset of
$\mathcal A_0$ that contains $\theta^*$ an \emph{admissible class}.
For an admissible class $\mathcal A$,
the \emph{covariance fiber} of $\Sigma^*$ and its \emph{minimum-edge fiber}
are
\[
 \E(\Sigma^*)=\{\theta\in\mathcal A:\Phi(\theta)=\Sigma^*\},\qquad
 \E_{\min}(\Sigma^*)=\argmin_{\theta\in\E(\Sigma^*)}c(\theta).
\]
Members of $\E(\Sigma^*)$ are the exact representations of $\Sigma^*$,
and members of $\E_{\min}(\Sigma^*)$ are its minimum-edge representations.
\end{definition}

For the unrestricted class, the two curves that form the
covariance fiber of $\Sigma_r$ in Example~\ref{ex:two_variable} meet only at $(0,\Sigma_r)$, the representation with $t=0$ and a single bidirected edge, and the minimum-edge fiber consists of the three
one-edge representations, whose graphs are $1\to2$, $2\to1$ and
$1\leftrightarrow2$. The covariance fiber is also unbounded. As $t$
grows, the coefficient $t$ and the error variance $1-2rt+t^2$ grow
without limit, so the fiber contains representations of arbitrarily
large norm. Several of our arguments need the representations to stay
in a bounded region, so in Section~\ref{sec:estimation} we restrict the
coefficients and the eigenvalues of the error covariance to fixed
intervals; the resulting domain $\K$ is closed and bounded, that is,
compact.

Since in the same example $\Sigma_r$ does not determine which
representation generated it, we take the whole minimum-edge fiber as the
target, and we measure the error of an estimate $\widehat\theta$ by
its distance to the nearest member of that set,
\[
 \dist(\widehat\theta,\E_{\min})
 =\inf_{\theta\in\E_{\min}}\|\widehat\theta-\theta\|.
\]
An estimator is therefore consistent when this distance tends
to zero in probability. The estimate need not settle near any single
minimum-edge representation; it may move among them as $n$ grows.

\subsection{Algebraic and Markov equivalence}\label{sec:equivalence}
To compare graphs, we look at the covariance matrices each
graph can produce. For an ADMG $H$, the \emph{covariance model}
$\mathcal M_H$ is the set of all matrices $\Phi(B,\Omega)$ in
\eqref{eq:model} for which $B_{ij}$ is nonzero only if $H$ contains
$i\to j$, the off-diagonal entry $\Omega_{ij}$ is nonzero only if $H$
contains $i\leftrightarrow j$, and $\Omega$ is positive definite. Unlike the covariance fiber $\E(\Sigma^*)$, no covariance is fixed here, and $\mathcal M_H$ collects the covariances of all representations whose support graph
lies in $H$. The \emph{free coordinates} of $H$ are its $p$
error variances and one coordinate for each of its $e(H)$ edges. We write
$\Phi_H$ for the covariance map as a function of these coordinates and
$\Theta_H\subseteq\mathbb R^{p+e(H)}$ for the open set of coordinates at
which $\Omega$ is positive definite, so that
$\mathcal M_H=\Phi_H(\Theta_H)$. Because an entry that $H$ allows may still be zero,
$\mathcal M_H$ also contains the covariance model of every subgraph of
$H$. For example, the graph with the single edge $1\to2$ produces every
positive definite $2\times2$ matrix $\Sigma$. To see this, take $\Omega_{11}=\sigma_{11}$,
the regression coefficient $B_{12}=\sigma_{12}/\sigma_{11}$ of $X_2$ on
$X_1$, and the residual variance
$\Omega_{22}=\sigma_{22}-\sigma_{12}^2/\sigma_{11}$, which is positive
because $\Sigma$ is positive definite.

A graph constrains its covariance model through polynomial equations
in the entries $\sigma_{ij}$ of $\Sigma$. For the chain $1\to2\to3$, for
instance, every covariance in the model satisfies
$\sigma_{13}\sigma_{22}-\sigma_{12}\sigma_{23}=0$, whatever the values of
the coefficients and error variances, because $X_1$ and $X_3$ are
independent given $X_2$. Such an equation holds throughout
$\mathcal M_H$, that is, at every covariance the graph can produce, so
it is a property of the graph alone. In algebraic geometry, the set of
points at which a collection of polynomials vanishes is called a
\emph{variety}, and the smallest variety that contains a given set is its
\emph{Zariski closure} \citep{sullivant2018}. The next
definition takes all polynomial equations of $\mathcal M_H$ together.

\begin{definition}[Covariance variety and algebraic equivalence]\label{def:variety}
For an ADMG $H$, the \emph{covariance variety} $\V_H$ is the set of
complex symmetric matrices satisfying every polynomial equality that holds
throughout $\mathcal M_H$,
\[
 \V_H=\bigl\{\Sigma\in\mathcal S_p(\mathbb C):g(\Sigma)=0\text{ for every
 polynomial }g\text{ with }g(\Sigma')=0\text{ for all }
 \Sigma'\in\mathcal M_H\bigr\},
\]
where $\mathcal S_p(\mathbb C)$ denotes the complex symmetric $p\times p$
matrices; equivalently, it is the smallest algebraic set
containing the model, its complex Zariski closure. Two graphs $G$ and $H$
are \emph{algebraically equivalent} when $\V_G=\V_H$.
\end{definition}

For the chain $1\to2\to3$, the covariance variety is the set
of all complex symmetric $3\times3$ matrices with
$\sigma_{13}\sigma_{22}=\sigma_{12}\sigma_{23}$. It contains
$\mathcal M_H$ and is larger, because it keeps the equations of the
model and drops its inequalities. It contains, for instance, real
symmetric matrices that satisfy the equation but are not positive
definite. On two vertices, a
nonempty graph satisfies no polynomial equation, so its covariance
variety is the set of all complex symmetric $2\times2$ matrices, and the support
graphs of all representations in Example~\ref{ex:two_variable} are
algebraically equivalent. We allow complex entries because the results
on dimension that we use in Section~\ref{sec:identification} are stated
for complex varieties. This does not change the variety, because an equation that
holds for all real parameter values with positive definite $\Omega$ also
holds for all complex parameter values, so the real model and its
complex extension have the same equations (Section~S1 of the Supplementary Material,
\citealp{supplement}). See
\citet{drton} and \citet{sullivant2018} for an introduction to the algebraic approach to structural
equation models.

Algebraic equivalence need not imply
$\mathcal M_G=\mathcal M_H$, particularly when bows are allowed. For example,
the instrumental-variable graph with edges $1\to2$, $2\to3$ and
$2\leftrightarrow3$ and the complete acyclic graph with edges $1\to2$,
$2\to3$ and $1\to3$ both have all symmetric $3\times3$ matrices as their
variety. With $\sigma_{ij}$ the covariance entries, however, the first model
satisfies $\sigma_{13}=B_{23}\sigma_{12}$
and so contains no covariance with $\sigma_{12}=0\ne\sigma_{13}$, while
the second contains every positive definite matrix.

\begin{figure}[t]
\centering
\includegraphics[width=\linewidth]{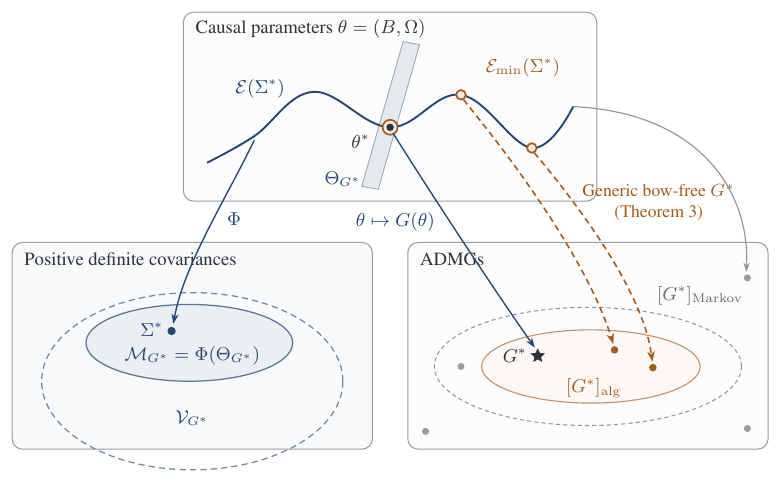}
\caption{\textbf{Parameters, Covariances and Graphs.} The
covariance map $\Phi$ sends every point of the covariance fiber
$\E(\Sigma^*)$ to $\Sigma^*$, and the support map $\theta\mapsto
G(\theta)$ sends each representation to its graph. The parameters
supported in $G^*$ form $\Theta_{G^*}$, whose image is the covariance
model $\mathcal M_{G^*}$, and the covariance variety $\V_{G^*}$ keeps the
equations of the model and drops its inequalities. Orange circles mark
minimum-edge representations. For a generic parameter of a bow-free
$G^*$, the slice $\Theta_{G^*}$ meets the fiber only at $\theta^*$, since
bow-free models are generically identifiable \citep{britopearl2002}, and
$G^*\in\{G(\theta):\theta\in\E_{\min}(\Sigma^*)\}
\subseteq[G^*]_{\mathrm{alg}}\subseteq[G^*]_{\mathrm{Markov}}$
by Theorem~\ref{thm:ident} and Lemma~\ref{lem:ci}. A representation with
more edges can have a graph outside the Markov class (gray arrow).}
\label{fig:spaces}
\end{figure}

Two ADMGs are Markov equivalent when they imply the same conditional independences, which are read off by m-separation \citep{richardson,richardson2003}. This criterion applies the rule of d-separation to paths that may contain bidirected edges, counting an internal vertex as a collider when both incident edges have arrowheads at it. For
Gaussian models, $X_i\perp X_j\mid X_S$ is equivalent to
the vanishing of the minor
$\det\Sigma_{\{i\}\cup S,\{j\}\cup S}$ of the covariance, a polynomial in
its entries,
and this minor vanishes throughout $\mathcal M_H$ exactly when $i$ and $j$
are m-separated given $S$ in $H$, bows included \citep{spirtes1998}.
Consequently, equality of covariance varieties implies equality
of graphical independence constraints, although the varieties can also
encode equalities that are not conditional independences, such as the
rank constraints characterized by trek separation \citep{std2010}.
Lemma~\ref{lem:ci} in Section~\ref{sec:identification} shows
that two graphs with the same covariance variety are Markov
equivalent. We call the set of graphs $H$ with $\V_H=\V_{G^*}$ the
\emph{generating algebraic class} $[G^*]_{\mathrm{alg}}$, or simply the \emph{generating class},
and the set of graphs Markov equivalent to $G^*$ the \emph{generating
Markov class} $[G^*]_{\mathrm{Markov}}$; by Lemma~\ref{lem:ci}, the first is contained in the
second. Figure~\ref{fig:spaces} shows how these objects fit together.

We can now state the question of this paper. Classical identifiability
concerns whether the population covariance $\Sigma^*$ determines the
generating parameter $\theta^*$. Example~\ref{ex:two_variable} shows
that it does not when the graph is unknown, so we pose a weaker question.
Does the rule of choosing a sparsest representation recover the
generating graph up to equivalence? That is, does every minimum-edge
representation of $\Sigma^*$ have a graph with the same covariance variety
as the generating graph, and hence, by Lemma~\ref{lem:ci}, the same
conditional independences? Here a graph's independences are
those it implies for every covariance it can produce. All
representations of $\Sigma^*$ share the independences that hold in
$\Sigma^*$, but their graphs can imply different ones; a complete
graph, for instance, represents every covariance and implies no
independence. This is a question about $\Sigma^*$ alone. The
second question is statistical. It concerns whether global minimizers of
a penalized likelihood find such representations from a sample, and
whether alternating updates reach them.

\begin{problem}
Find conditions on the generating graph and its parameter under
which
\[
 \theta\in\E_{\min}(\Sigma^*)
 \quad\Longrightarrow\quad
 \V_{G(\theta)}=\V_{G(\theta^*)},
\]
and decide whether global minimizers of a penalized likelihood
recover members of $\E_{\min}(\Sigma^*)$ from a sample, and whether
alternating updates find them.
\end{problem}

Section~\ref{sec:identification} answers the first part and
Sections~\ref{sec:estimation}--\ref{sec:computation} the second. The
implication goes in one direction only. Every minimum-edge representation has a graph in the generating algebraic class, whereas some graphs in that class may fail to represent $\Sigma^*$ exactly (as in the instrumental-variable example of Section~\ref{sec:equivalence}) or may fall outside the chosen parameter domain.
\section{Identification by joint sparsity}\label{sec:identification}
This section answers the first part of the Problem. It shows that, for a
generic parameter of a bow-free generating graph, the graph of every
minimum-edge representation lies in the generating class. The proof
compares the dimensions of covariance models, where the dimension of
$\V_H$ is, roughly, the number of independent directions in which the
graph $H$ can move the covariance (Section~\ref{sec:dimension}). A bow-free graph has dimension equal to its number of free parameters, whereas any ADMG has dimension at most its number of free parameters (Lemma~\ref{lem:dimension}), so a graph
with fewer edges than the generating graph cannot produce all of its
covariances. Theorem~\ref{thm:ident} turns this into the identification
result, Lemma~\ref{lem:ci} adds Markov equivalence, and
Corollary~\ref{cor:regular} shows that the minimum-edge fiber is
generically finite.

The result cannot hold for every parameter value. In the complete graph
with edges $1\to2$, $2\to3$ and $1\to3$ and unit error variances, for
example, $\sigma_{13}=B_{13}+B_{12}B_{23}$, so the parameters with
$B_{13}=-B_{12}B_{23}$ make $X_1$ and $X_3$ uncorrelated, and the
covariance then has a representation with fewer edges. Exceptions of
this kind are parameter values at which a polynomial in the parameters
vanishes, and we exclude them.

\begin{definition}[Generic property]\label{def:generic}
A property of the parameters of a graph $H$ holds
\emph{generically} when the parameter values at which it fails all
satisfy $f=0$ for a fixed nonzero polynomial $f$ in the free coordinates
of $H$; that is, there is a polynomial $f\not\equiv0$ on
$\mathbb R^{p+e(H)}$ such that the property holds at every
$\vartheta\in\Theta_H$ with $f(\vartheta)\ne0$. We call the zero set
$\{\vartheta:f(\vartheta)=0\}$ an \emph{exceptional set}; it is a proper
algebraic subset of $\mathbb R^{p+e(H)}$ and has Lebesgue measure zero.
For a representation $\theta$, we write $f(\theta)$ for the value of $f$
at the free coordinates of $\theta$.
\end{definition}

\begin{assumption}[Generating model]\label{ass:generating}
The generating graph $G^*$ is a bow-free ADMG with $e^*$
edges, and its parameter $\theta^*$, with free coordinates
$\vartheta^*\in\Theta_{G^*}\subseteq\mathbb R^{p+e^*}$, is generic, that is, there
is a polynomial $f\not\equiv0$ on $\mathbb R^{p+e^*}$, depending only on
$G^*$, with $f(\vartheta^*)\ne0$. Each result below that uses this
assumption exhibits such a polynomial, and $\vartheta^*$ lies outside the
exceptional sets of all of them.
\end{assumption}

\begin{remark}[Scope of genericity]\label{rem:genericity}
For a fixed graph, a proper polynomial zero set has Lebesgue measure
zero, so the generic conclusions hold almost surely under a parameter
distribution with a density in the free coordinates. The
assumption places no distribution over graphs, and since it imposes no
uniform distance from the exceptional values, a generic parameter may be close to one at which
a competing model becomes observationally indistinguishable. Genericity
is therefore a pointwise notion that removes only a Lebesgue-null set.
Strong faithfulness, under which the PC algorithm is uniformly
consistent, removes a set
of positive measure, which \citet{uhler2013} show can be large even for
small sparse graphs.
\end{remark}

\subsection{Dimension and equivalence}\label{sec:dimension}
Our identification proof rests on two properties of the
covariance variety $\V_H$, its dimension and its irreducibility. Recall
from Section~\ref{sec:equivalence} that a graph $H$ has $p+e(H)$ free
coordinates, the $p$ error variances and one coordinate for each of its
$e(H)$ edges, and that $\Phi_H$ sends them to a covariance matrix. The
dimension of $\V_H$ is the number of independent directions in which
small changes of these coordinates move the covariance, that is, the
rank of the Jacobian of $\Phi_H$ at a generic point of $\Theta_H$. It is
at most $p+e(H)$, and it is smaller when two coordinates move the
covariance in the same direction, as the two edges of a bow can
(Example~\ref{ex:dimension}). The second property is irreducibility. A variety is \emph{irreducible} if it is not the union
of two smaller varieties. Every smaller variety inside an irreducible
variety has strictly lower dimension, so if one irreducible variety
contains another of the same dimension, the two are equal. This is how
Theorem~\ref{thm:ident} turns a containment between two covariance
models into their equality.

\begin{lemma}\label{lem:dimension}
For every acyclic directed mixed graph $H$, $\V_H$ is irreducible and
$\dim\V_H\le p+e(H)$. If $H$ is bow-free, equality holds.
\end{lemma}

The equality for bow-free graphs is easiest to see at
$B=0$ and $\Omega=I$. There a small change of an error variance moves
one diagonal entry of the covariance, and a small change of the
coefficient or error covariance on an edge between $i$ and $j$ moves the
entry $\sigma_{ij}$. In a bow-free graph no two edges join the same pair,
so every parameter moves a different entry, and the $p+e(H)$ directions
are independent. Section~S1 of the Supplementary Material
\citep{supplement} gives the proof.

\begin{example}[Dimension with and without a bow]\label{ex:dimension}
The graph with the single directed edge $1\to2$ has three
free parameters, and, as shown in Section~\ref{sec:equivalence}, its
covariance model is the set of all positive definite $2\times2$
matrices, which has dimension three, so Lemma~\ref{lem:dimension} holds
with equality. The bowed graph with
$1\to2$ and $1\leftrightarrow2$ has four free parameters but the
same covariance model, so $\dim\V_H=3<4$, because the extra parameter moves along
the fiber of Example~\ref{ex:two_variable}. Here the bow costs an edge
without adding a dimension. A bow need not lower the dimension (the bowed
instrumental-variable graph of Section~\ref{sec:equivalence} has full
dimension), but the dimension count below uses only the upper bound of
Lemma~\ref{lem:dimension} for competitors, so it can allow bows among competitors.
\end{example}

To pass from equal varieties to Markov equivalence, we use the
characterization of conditional independences by vanishing minors
recalled in Section~\ref{sec:equivalence}; Section~S2 of the
Supplementary Material proves the following lemma.

\begin{lemma}\label{lem:ci}
Equality $\V_G=\V_H$ implies Markov equivalence for acyclic directed mixed graphs, including graphs with bows.
\end{lemma}

We can now state the identification result. It concerns every
exact representation of $\Sigma^*$ whose cost is at most that of the
generating parameter, and it shows that such a representation has the
same cost and a graph in the generating class, unless $\theta^*$ lies in
the zero set of a polynomial $f$ determined by $G^*$.

\begin{theorem}[Generic equivalence-class identification]\label{thm:ident}
Fix a bow-free acyclic generating support $G^*$ with $e^*$ edges.
There is a nonzero polynomial $f$ in the $p+e^*$ free parameters of
$G^*$ such that, whenever $f(\theta^*)\ne0$, every
acyclic mixed representation $\theta$ satisfying
\[
 \Phi(\theta)=\Phi(\theta^*),\qquad c(\theta)\le e^*
\]
has $c(\theta)=e^*$ and $\V_{G(\theta)}=\V_{G^*}$. In particular, every
minimum-edge representation in any admissible class containing $\theta^*$ is
algebraically and Markov equivalent to $G^*$.
\end{theorem}

In words, for a generic parameter of a bow-free graph, no exact
representation of $\Sigma^*$ has fewer edges than $\theta^*$, and every
exact representation with as many edges has a graph that imposes the
same polynomial equalities on the covariance as $G^*$, and hence the
same conditional independences.

The polynomial $f$ rules out coincidences. A competing graph
$H$ with at most $e^*$ edges can represent $\Sigma^*$ in two ways. If its
variety contains $\V_{G^*}$, the dimension count below forces
$e(H)=e^*$ and $\V_H=\V_{G^*}$, which is the conclusion of the theorem.
If its variety does not contain $\V_{G^*}$, the two varieties meet in a
lower-dimensional subset of $\V_{G^*}$, and $\Sigma^*$ falls into it only
for special parameter values, such as $B_{13}=-B_{12}B_{23}$ in the
example at the start of this section. For each such $H$, these values
are the zeros of a nonzero polynomial in the parameters of $G^*$, and
$f$ is the product of these polynomials over the finitely many such
graphs. So $f(\theta^*)\ne0$ ensures that no graph with at most $e^*$ edges whose variety does not contain $\V_{G^*}$ can represent $\Sigma^*$ by accident.

The proof compares dimensions. When $f(\theta^*)\ne0$, the support
graph $H$ of an exact representation with at most $e^*$ edges must
satisfy $\V_{G^*}\subseteq\V_H$, whence, by Lemma~\ref{lem:dimension},
\[
 p+e^*=\dim\V_{G^*}\le\dim\V_H\le p+e(H)\le p+e^*.
\]
Every inequality is an equality, so $e(H)=e^*$, and irreducibility turns the
containment into $\V_H=\V_{G^*}$; Markov equivalence then follows
from Lemma~\ref{lem:ci}. This containment-and-dimension step is the
argument of \citet{dong2026}, who compare likelihood maximizers by model
dimension (Remark~\ref{rem:sparsest}); Section~S3 of the Supplementary
Material gives the full proof.

Since $\theta^*$ is itself an exact representation with $e^*$
edges, we conclude from Theorem~\ref{thm:ident} that, under
Assumption~\ref{ass:generating}, in every admissible class the
minimum edge count is $e^*$ and $\theta^*\in\E_{\min}(\Sigma^*)$; by
Corollary~\ref{cor:regular} below, this target is generically
a finite set of representations, for instance the three
one-edge representations of Example~\ref{ex:two_variable}, even though
the covariance fiber around it can be a continuum.

\begin{example}[Three vertices]\label{ex:three_vertex}
(a) Let $G^*$ be the chain $1\to2\to3$ with $B^*_{12}B^*_{23}\ne0$
and arbitrary error variances, and take the unrestricted class
$\mathcal A_0$. Since every off-diagonal entry of $\Sigma^*$ is nonzero,
the minimum edge count is two, and since
$X_1\perp X_3\mid X_2$ is the only conditional independence of $\Sigma^*$,
the two edges join the pairs $\{1,2\}$ and $\{2,3\}$ without a collider at
vertex~$2$. Each of the five such graphs, $1\to2\to3$,
$1\leftarrow2\leftarrow3$, $1\leftarrow2\to3$, $1\leftrightarrow2\to3$ and
$1\leftarrow2\leftrightarrow3$, carries exactly one exact representation,
so $\E_{\min}(\Sigma^*)$ has five members, two of them with a bidirected
edge, and all five graphs have the same variety, the hypersurface
$\sigma_{13}\sigma_{22}=\sigma_{12}\sigma_{23}$. Exact representations
with a bow need at least three edges and can form continua, for instance
on the graph with edges $1\to2$, $1\leftrightarrow2$ and $2\to3$, so,
as in Example~\ref{ex:two_variable}, a finite minimum-edge fiber sits
inside an infinite covariance fiber.

(b) Let $G^*$ be the complete DAG $1\to2\to3$, $1\to3$ with
$B^*_{12}=a$, $B^*_{23}=b$, $B^*_{13}=c$ and error variances $s,t,u$, and
suppose that $abc(ab+c)\sigma^*_{23}(au-bct)\ne0$. Then every exact
representation in $\mathcal A_0$ has at least three edges, and the
instrumental-variable graph of Section~\ref{sec:equivalence}, with edges
$1\to2$, $2\to3$ and $2\leftrightarrow3$, carries exactly one exact
representation, with $B_{23}=b+c/a$ and $\Omega_{23}=-ct/a\ne0$. This
representation belongs to $\E_{\min}(\Sigma^*)$ and contains a bow,
although $G^*$ is bow-free; in agreement with Theorem~\ref{thm:ident}, its
graph has the same variety as $G^*$, the space of all symmetric $3\times3$
matrices.

Section~S3 of the Supplementary Material checks the
calculations in both parts. The chain of part (a) is also one of the
three generating graphs in the numerical study of
Section~\ref{sec:global}.
\end{example}

\begin{remark}[Generating graphs with bows]\label{rem:bows}
We use bow-freeness of $G^*$ only through the equality
$\dim\V_{G^*}=p+e^*$, which
is the generic identifiability of bow-free models
\citep{britopearl2002,fdd2012} expressed as a dimension count; the
competing graphs enter only through the upper bound of
Lemma~\ref{lem:dimension} and may contain bows. Theorem~\ref{thm:ident}
and Corollary~\ref{cor:regular} therefore hold verbatim with ``bow-free''
replaced by $\dim\V_{G^*}=p+e^*$, and so do the later results that use
Assumption~\ref{ass:generating} only through them (Section~S3 of the
Supplementary Material). This condition holds exactly when,
for a generic parameter of $G^*$, only finitely many parameter values of
$G^*$ produce the same covariance (the covariance map of $G^*$ is
generically finite-to-one). It therefore holds for every generically
identifiable graph, for which that number is one, for instance every graph that satisfies the half-trek criterion of
\citet{fdd2012}. The instrumental-variable graph of
Example~\ref{ex:three_vertex}(b) contains a bow and is generically
identifiable, since its six parameters are recovered from the covariance,
starting from $B_{23}=\sigma_{13}/\sigma_{12}$. Without the
condition the conclusion can fail. For the bowed graph of
Example~\ref{ex:dimension}, every covariance has a one-edge
representation, so a generating parameter with both edges nonzero is not
a minimum-edge representation.
\end{remark}

\begin{remark}[Why the two edge types are counted equally]\label{rem:weights}
By Lemma~\ref{lem:dimension}, a bow-free graph $H$ has
$e(H)=\dim\V_H-p$, so each directed and each bidirected edge adds one
dimension, and when we compare dimensions in the proof of
Theorem~\ref{thm:ident} we charge the two edge types equally. With unequal
charges, the cheapest exact representations can leave the generating
class. Give each directed edge weight one and each bidirected edge weight
$w\in(0,1/2)$, and let $G^*$ be the chain $1\to2\to3$ with
$B^*_{12}B^*_{23}\ne0$. Every graph in the algebraic class of $G^*$ has at
least two edges, at least one of them directed, so it costs at least
$1+w$, attained by $1\leftrightarrow2\to3$. The exact
representation $(0,\Sigma^*)$ on the complete bidirected graph costs only
$3w<1+w$ and is neither algebraically nor Markov equivalent to $G^*$.
Every minimizer of the weighted count over any admissible class containing
$(0,\Sigma^*)$, such as $\mathcal A_0$, therefore lies outside the
generating algebraic class, and the weighted analog of
Theorem~\ref{thm:ident} fails (Section~S3 of the Supplementary Material).
\end{remark}

\begin{remark}[Comparison with sparsest-representation results]\label{rem:sparsest}
For directed acyclic graphs, \citet{vandegeer} target the minimal-edge
independence map through an $\ell_0$-penalized likelihood, and
\citet{raskutti} formalize the sparsest Markov representation assumption
that underlies permutation-based search, whose greedy versions
\citet{solus2021} prove consistent. With latent variables,
\citet{bernstein2020} show that a sparsest independence map lies in the
Markov equivalence class of the true ancestral graph under conditions
weaker than faithfulness. Also, \citet{dcd} show that, when the
distribution is Markov and faithful to an arid generating graph, the
global minimizer of the Bayesian information criterion (BIC) over a class of
mixed graphs asymptotically implies the same equality constraints as that
graph, which include those of the nested Markov property
\citep{nested2023}. For partially observed linear Gaussian causal models,
\citet{dong2026} show under a generalized faithfulness condition that a
graph maximizing the likelihood with the smallest model dimension among
the maximizers is algebraically equivalent to the true graph. Their proof
first shows that the model of the true graph is contained in the model of
the selected graph and then compares dimensions, which is also the
argument of Theorem~\ref{thm:ident}, and \citet{ommen2024} had already
used the edge count of a bow-free graph as its model dimension to reduce
equivalence to such a containment. Like \citet{bernstein2020} and
\citet{dcd}, we count directed and bidirected edges together.
Theorem~\ref{thm:ident} differs from those latent-variable results in three ways. First,
the generating graph need only be bow-free, a weaker requirement than the
arid graphs of \citet{dcd}, or more generally generically identifiable
(Remark~\ref{rem:bows}). Second, the competing graphs may be any acyclic
mixed graphs, including graphs with bows, which can represent a covariance
by a continuum of parameters. Third, genericity of the generating
parameter replaces faithfulness. Unlike \citet{bernstein2020}, we assume a
linear Gaussian model, and we conclude algebraic, and hence Markov,
equivalence. The main difficulty is that the edge count of a bow-free
graph determines its model dimension (Lemma~\ref{lem:dimension}), while
for a graph with bows it only bounds the dimension from above, so the
proof must show that minimizing the edge count still selects the right
class when competitors with bows are allowed.
\end{remark}

\subsection{Regularity of the minimum-edge fiber}\label{sec:regularity}
Once a minimum-edge support is fixed, we need the parameters on
that support to be locally identifiable, because
a correctly selected graph class need not
provide a nonsingular estimation problem.

\begin{definition}[Regular representation]\label{def:regular}
A representation $\theta$ with support graph $H$ is
\emph{regular on its support} when the covariance Jacobian with respect to
the free coordinates of that support has full column rank,
that is, $\operatorname{rank}D\Phi_H(\vartheta)=p+e(H)$ at the free
coordinates $\vartheta\in\Theta_H$ of $\theta$; equivalently, the
Gaussian Fisher information on the support is positive definite.
\end{definition}

Under the conditions of Theorem~\ref{thm:ident}, every minimum-edge
representation has this property unless $\theta^*$ lies in a
further exceptional set.

\begin{corollary}[Regular minimum-edge fibers]\label{cor:regular}
Under the conditions of Theorem~\ref{thm:ident}, there is a
further nonzero polynomial $g$ in the free coordinates of $G^*$ such
that, whenever $f(\theta^*)g(\theta^*)\ne0$, the minimum-edge fiber
$\E_{\min}(\Sigma^*)$ of the unrestricted class $\mathcal A_0$ is finite
and every member is regular on its support. Moreover, for any constants
$M_B>0$ and $0<m<M_O$ fixed in advance, there is a nonzero polynomial
$h$, depending also on these constants, such that, whenever in addition
$h(\theta^*)\ne0$, every $(B,\Omega)\in\E_{\min}(\Sigma^*)$ with
$\max_{i,j}|B_{ij}|\le M_B$ and $mI\preceq\Omega\preceq M_OI$ satisfies
$\max_{i,j}|B_{ij}|<M_B$ and $mI\prec\Omega\prec M_OI$.
\end{corollary}

The second part says that, generically, no minimum-edge
representation sits exactly on the boundary of the region defined by these bounds, with a
coefficient equal to $\pm M_B$ or an eigenvalue of $\Omega$ equal to $m$
or $M_O$. Lying on the boundary is an equation in the parameters, so it
happens only on an exceptional set. Interior points matter in
Sections~\ref{sec:estimation} and~\ref{sec:computation}, where we perturb
a representation in every direction without leaving the domain. The
bounds are fixed in advance because the polynomial $h$ depends on them.
Section~S4 of the Supplementary Material gives the proof.
\section{Penalized estimation}
\label{sec:estimation}
Minimizing the joint edge count requires the exact
representations of $\Sigma^*$, which are unknown in practice. In this section we replace this exact edge-count criterion by a penalized likelihood that can be computed from data. The Gaussian likelihood measures how well a representation fits the
covariance, and a penalty measures how many edges it uses. The
likelihood alone cannot select a sparsest representation, because all
exact representations fit $\Sigma^*$ equally well, including those with
redundant edges. We therefore add a continuous penalty that equals one
for every coefficient or error covariance entry above a threshold and is
smaller for smaller entries, and we show that the minimizers of the
penalized population likelihood are exactly the minimum-edge
representations. The difficulty is that one threshold must work for
every representation in a covariance fiber, which can be a continuum.

Our arguments compare representations across the whole fiber, so the
fiber must be bounded, which it need not be in the unrestricted class
(Section~\ref{sec:fiber}). We therefore restrict the parameters to a
compact domain.

\begin{assumption}[Compact parameter domain]\label{ass:domain}
Fix $0<m<M_O<\infty$ and $M_B>0$, and let $\K$ consist of all acyclic
$B$ with $|B_{ij}|\le M_B$ and all $mI\preceq\Omega\preceq M_OI$.
The generating parameter belongs to $\K$.
\end{assumption}

From now on we impose Assumption~\ref{ass:domain} and take
$\K$ as the admissible class. The covariance fiber $\E$ and the
minimum-edge fiber $\E_{\min}$ are defined relative to $\K$, and
$e=\min_{\theta\in\E}c(\theta)$ is the smallest cost of an exact
representation in $\K$. Both $\E$ and $\E_{\min}$ are compact. The fiber
$\E$ is closed and bounded, because it is the set of points of the
compact $\K$ at which the continuous map $\Phi$ equals $\Sigma^*$. The set
$\E_{\min}$ is closed as well. If exact representations with $e$ edges
converge, a coordinate that is nonzero at the limit is nonzero along the
tail of the sequence, so the limit has at most $e$ edges, and it has at
least $e$ by the definition of $e$. Under Assumption~\ref{ass:generating},
Theorem~\ref{thm:ident} gives $e=e^*$, so $\theta^*\in\E_{\min}$. The domain
$\K$ bounds the entries of $B$ and the eigenvalues of $\Omega$ but not
their supports, so representations with bows remain allowed. We call
$mI\preceq\Omega\preceq M_OI$ the \emph{eigenvalue constraints}; they
keep every error covariance in $\K$ positive definite and well
conditioned. Like $\E_{\min}$
and $e$, the tuning parameters below depend on the bounds.

For a covariance input $S$, the Gaussian
negative log likelihood of a representation $\theta=(B,\Omega)$ is, per
observation and up to an additive constant,

\begin{equation}\label{eq:loss}
 \ell_S(\theta)=\tfrac12\log\det\Omega+
 \tfrac12\operatorname{tr}\{\Omega^{-1}(I-B)^T S(I-B)\}.
\end{equation}
The usual term $\log\det(I-B)$ does not appear because
acyclicity implies $\det(I-B)=1$.
We set $\ell=\ell_{\Sigma^*}$ and $\ell_*=\min_\K\ell$, so that
$\ell(\theta)-\ell_*$ is the Kullback--Leibler divergence from
$N(0,\Sigma^*)$ to $N(0,\Phi(\theta))$ and vanishes precisely on $\E$.

\subsection{Population minimizers}
\label{sec:population}
Recall from Corollary~\ref{cor:regular} that, under its generic
conditions, the minimum-edge fiber is finite; the covariance fiber that
surrounds it is not.
For any $B$ compatible with a fixed topological order, the error covariance
$\Omega_{\Sigma^*}(B)=(I-B)^T\Sigma^*(I-B)$
produces the same covariance $\Sigma^*$, and small changes in $B$ preserve
feasibility around an interior representation. For $p\ge2$, we therefore
obtain a continuous family of exact representations, as in
Example~\ref{ex:two_variable}. A penalty that selects the
minimum-edge fiber must compare its members with all of these
representations, including those whose additional edges are arbitrarily
small.

Let $a(\theta)$ list the directed-edge coordinates and the off-diagonal
entries of $\Omega$, with one entry for each unordered pair.
We use a rescaling of the minimax concave penalty of
\citet{mcp2010} whose maximum value is one, which we call the
\emph{quasi-MCP penalty},
\begin{equation}\label{eq:penalty}
 q_\delta(t)=
 \begin{cases}2|t|/\delta-t^2/\delta^2,&|t|\le\delta,\\
 1,&|t|\ge\delta,
 \end{cases}
 \qquad P_\delta(\theta)=\sum_kq_\delta(a_k(\theta)).
\end{equation}
The penalty rises continuously from zero to one and is constant for
$|t|\ge\delta$. We call a penalty that is constant beyond a threshold
\emph{saturated}, and we
say that the penalty of a coordinate $a_k$ is saturated when
$|a_k|\ge\delta$. Our penalty differs from the quasi-MCP
penalty of \citet{deng} only in its scaling. Theirs equals
$(\lambda\delta/2)q_\delta$ in our notation, so the price of a large edge,
$\lambda\delta/2$, changes with the threshold. Because we scale
$q_\delta$ to have maximum value one, every edge of magnitude at least
$\delta$ adds exactly $\lambda$ to $\ell+\lambda P_\delta$, and the
threshold $\delta$ and the price $\lambda$ of an edge can be chosen
separately.
Figure~\ref{fig:fiber}\textbf{B} in Section~\ref{sec:l1}
illustrates this flat portion along a covariance fiber.

\begin{theorem}[Population selection by saturation]\label{thm:penalty}
There exist $\delta_0,\lambda_0>0$ such that, for
$0<\delta\le\delta_0$ and $0<\lambda<\lambda_0$,
\[
 \argmin_{\theta\in\K}\{\ell(\theta)+\lambda P_\delta(\theta)\}
 =\E_{\min}.
\]
\end{theorem}

Informally, when every edge of magnitude at least a small threshold
$\delta$ pays the same small price $\lambda$, the penalized population
likelihood is minimized over $\K$ exactly at the minimum-edge
representations, even though the covariance fiber around them can be a
continuum.

Theorem~\ref{thm:penalty} connects the identification result
to optimization. The penalized likelihood is not convex, yet every one of
its global minimizers is a minimum-edge representation, and, under
Assumption~\ref{ass:generating}, its graph lies in the generating class
(Theorem~\ref{thm:ident}). How to reach a global minimizer is a
separate question, which Section~\ref{sec:computation} addresses.

The proof chooses the threshold first and the price second. On the
compact fiber $\E$, the $e$th largest edge magnitude of an exact
representation has a positive minimum, since otherwise a limit of exact
representations would have fewer than $e$ edges. With $\delta$ below that
minimum, every parameter near $\E$ pays at least $e$ full penalties, so
none improves on the minimum-edge representations; parameters far from
$\E$ fit worse by a fixed amount, which a small $\lambda$ cannot offset.
The argument uses only the $e$ largest edge coordinates of each representation,
because the smallest nonzero edges on the fiber can be arbitrarily small;
Remark~\ref{rem:finite_fiber} returns to this point. Section~S5 of the
Supplementary Material \citep{supplement} gives the full proof.

\begin{remark}[Comparison with finite-fiber selection]\label{rem:finite_fiber}
\citet{deng} study acyclic structural equation models with
independent errors. In the linear Gaussian case the error variances are
arbitrary, so the graph is not identifiable from the covariance, but the
covariance has at most $p!$ exact representations, one for each vertex
order; for general models they assume that only finitely many parameters
generate the distribution. Under this finiteness, they show that a
saturated penalty selects the minimum-edge structures. Correlated errors
make the problem harder, because the covariance fiber can be a continuum, as the
family above shows, and generically only the minimum-edge fiber is
finite. \citet{deng}
also note that finiteness can be weakened to a positive lower bound on the
smallest nonzero edge over all exact representations; on the continuous
family this bound is zero, which is why we work with the $e$ largest edge
coordinates and with compactness of the domain.
\end{remark}

In the proof of Theorem~\ref{thm:penalty} we use neither the
quadratic form of quasi-MCP below its threshold nor its concavity; besides
continuity, we need only that every nonzero edge coordinate pays a positive
cost and every coordinate of magnitude at least $\delta$ pays exactly one unit.
\citet{deng} note that MCP or SCAD can replace quasi-MCP in
their setting; the next corollary shows that any continuous penalty with
these two properties works here, whether or not it is concave or
monotone. Its second part refers to Theorem~\ref{thm:sample} of
Section~\ref{sec:sampling}, which extends Theorem~\ref{thm:penalty} to a
sample covariance.

\begin{corollary}[Saturated penalty family]\label{cor:family}
Replace $q_\delta(t)$ by $\rho(|t|/\delta)$, where
$\rho:[0,\infty)\to[0,1]$ is continuous, $\rho(0)=0$,
$\rho(s)>0$ for $s>0$, and $\rho(s)=1$ for $s\ge1$.
Theorem~\ref{thm:penalty} and parts~\textup{(i)} and~\textup{(ii)} of
Theorem~\ref{thm:sample} remain valid.
\end{corollary}

Normalized MCP \citep{mcp2010}, normalized SCAD \citep{scad2001},
truncated $\ell_1$ \citep{tlp2012} and the capped powers
$\rho(s)=\min(s^r,1)$ for $r>0$ satisfy these conditions once their
plateaus are normalized to one; neither concavity nor monotonicity
is required.
Different plateau heights yield a weighted edge count, for which the
weighted analog of Theorem~\ref{thm:ident} can fail
(Remark~\ref{rem:weights}), and a penalty that approaches its
plateau only asymptotically lies outside the stated family.
The slope near zero, which saturation leaves free, determines whether a
sample minimizer removes an edge exactly or requires a subsequent
thresholding step. Section~S6 of the Supplementary Material
gives the proof of Corollary~\ref{cor:family}.

\subsection{Sampling, tuning and support recovery}
\label{sec:sampling}
With fixed tuning parameters, we first replace $\Sigma^*$ by a sample covariance. Let $S_n$ be the centered empirical covariance of independent observations from $N(0,\Sigma^*)$, normalized by $1/n$, and set
$F_n=\ell_{S_n}+\lambda P_\delta$ and $F=\ell+\lambda P_\delta$. Since
$\ell_S$ depends on $S$ only through a trace that is linear
in $S$, we have,
for fixed $p$,
\begin{equation}\label{eq:uniform}
 \sup_\K|F_n-F|\le C_\K\|S_n-\Sigma^*\|_F=O_p(n^{-1/2}),\quad
 C_\K=\tfrac12\sup_\K\|(I-B)\Omega^{-1}(I-B)^T\|_F,
\end{equation}
where $C_\K$ is finite by the parameter bounds; the comparison holds
over all supports and all representations of the same covariance.
Because a nonconvex criterion may be minimized only approximately, we
allow a nonnegative objective error $\eta_n$.

\begin{definition}[Approximate global minimizer; raw support]\label{def:approx}
A feasible estimate $\widehat\theta_n$ is an
\emph{$\eta_n$-approximate global minimizer} of $F_n$ when
\[
 F_n(\widehat\theta_n)\le\inf_{\theta\in\K}F_n(\theta)+\eta_n.
\]
Its \emph{raw support} is the set of nonzero edge coordinates before
thresholding; its thresholded support is the graph obtained by deleting the
edge coordinates below a stated threshold.
\end{definition}

We say that the $\eta_n$-approximate global minimizers converge in
distance to $\E_{\min}$ in probability when
$\sup\dist(\widehat\theta_n,\E_{\min})\to_p0$, the supremum being taken
over all of them. The raw and thresholded supports can differ,
because an arbitrarily small extra coordinate can change the support
graph while changing a continuous objective by less than $\eta_n$.

\begin{theorem}[Consistency of global minima]\label{thm:sample}
Fix tuning parameters satisfying Theorem~\ref{thm:penalty}.

\textup{(i)} All global minimizers of $F_n$ converge in distance to
$\E_{\min}$ in probability.

\textup{(ii)} This conclusion also holds for all
$\eta_n$-approximate global minimizers if $\eta_n\to_p0$.

\textup{(iii)} If every member $(B,\Omega)$ of $\E_{\min}$
satisfies $mI\prec\Omega\prec M_OI$, then, with probability tending to one, every exact
global minimizer has the same support as some member of $\E_{\min}$.
\end{theorem}

Parts~(i) and~(ii) rest on the uniform bound in \eqref{eq:uniform},
which transfers the population objective gap outside any neighborhood of
$\E_{\min}$ to the sample criterion. For part~(iii), we compare an exact
minimizer near a minimum-edge representation with the
parameter obtained by deleting a small additional edge. Since the
population likelihood derivative vanishes at an exact representation along
every coordinate that keeps the support acyclic, the quasi-MCP penalty
saved by the deletion exceeds the accompanying change in likelihood, so an
exact minimum cannot retain such an edge. Shrinking a coefficient
toward zero preserves acyclicity and the coefficient bound, so we
impose no condition on $M_B$ and need the strict inequalities of part~(iii) only to keep the error
covariance within the eigenvalue constraints. For approximate minimizers we remove
the small additional coordinates by thresholding. Section~S7 of
the Supplementary Material gives the proof.

On a recovered regular support, all active penalties are constant in a
neighborhood of the corresponding minimum-edge representation, so
the local estimation equations are those of the restricted unpenalized
likelihood, whose information matrix is positive definite by
Corollary~\ref{cor:regular}, which gives the following rate.

\begin{corollary}[Generic parametric rate]\label{cor:rate}
Under the generic conditions of Theorem~\ref{thm:ident} and
Corollary~\ref{cor:regular}, with the admissible fixed tuning parameters of
Theorem~\ref{thm:sample}, every exact sample global minimizer has distance
$O_p(n^{-1/2})$ to $\E_{\min}$.
\end{corollary}

The rate comes from the usual likelihood expansion, applied separately
near each of the finitely many regular minimum-edge
representations, so we control the distance to the whole fiber
without identifying which member the estimator approaches
(Section~S8 of the Supplementary Material).

The range of fixed tuning parameters allowed by Theorem~\ref{thm:penalty} depends on the unknown
$\Sigma^*$, so we let the tuning parameters shrink along a deterministic
schedule that eventually enters that range, as is standard in
the asymptotic theory of penalized likelihood \citep{scad2001}.

\begin{assumption}[Gaussian sampling and shrinking tuning parameters]\label{ass:tuning}
The dimension $p$ is fixed, and $S_n$ is the centered empirical covariance
of independent Gaussian observations, normalized by $1/n$.
The positive sequences $\lambda_n$ and $\delta_n$ tend to zero,
$n\lambda_n$ tends to infinity, and $\eta_n=o_p(\lambda_n)$.
\end{assumption}

The uniform bound \eqref{eq:uniform} would require
$n^{1/2}\lambda_n\to\infty$; a sharper comparison allows the edge price $\lambda_n$ to be
of the order $\log(n)/n$ of the penalty in the BIC. If we optimize $\ell_{S_n}$ over all positive definite covariances, the reduction relative to any exact representation of $\Sigma^*$ is at most
\[
A_n=\tfrac12\{\operatorname{tr}(S_n\Sigma^{*-1})
-\log\det(S_n\Sigma^{*-1})-p\}=O_p(n^{-1}),
\]
since $2nA_n$ is the likelihood-ratio statistic for $\Sigma=\Sigma^*$,
with a limiting $\chi^2$ distribution on $p(p+1)/2$ degrees of freedom.
Comparing an approximate minimizer with a minimum-edge representation
therefore gives $P_{\delta_n}(\widehat\theta)\le e+(A_n+\eta_n)/\lambda_n$,
so the excess penalty tends to zero in probability whenever
$n\lambda_n\to\infty$ and $\eta_n=o_p(\lambda_n)$. We need a
lower bound on $\lambda_n/\delta_n$ only for support recovery by an exact
minimizer without thresholding.

\begin{theorem}[Shrinking penalties]\label{thm:shrinking}
Suppose Assumption~\ref{ass:tuning} holds.

\textup{(i)} All $\eta_n$-approximate global minima of
$\ell_{S_n}+\lambda_nP_{\delta_n}$ converge in distance to $\E_{\min}$
in probability.

\textup{(ii)} For each fixed $\tau>0$, thresholding their edge coordinates at $\tau\delta_n$ recovers
the support of a member of $\E_{\min}$ with probability tending to one.

\textup{(iii)} If every member $(B,\Omega)$ of $\E_{\min}$
satisfies $mI\prec\Omega\prec M_OI$
and $\liminf_n\lambda_n/\delta_n>0$, every exact global minimum has this
support property without thresholding.

\textup{(iv)} Under the generic conditions of Theorem~\ref{thm:ident} and
Corollary~\ref{cor:regular}, the exact minima
in part~\textup{(iii)} also have distance $O_p(n^{-1/2})$ to $\E_{\min}$.
\end{theorem}

The theorem says that, when the price of an edge tends to zero more slowly than
$1/n$, approximate minimizers with objective error $o_p(\lambda_n)$
approach the minimum-edge fiber, and deleting their edge coordinates
below a fixed fraction of $\delta_n$ recovers a minimum-edge support with
probability tending to one.

In the proof, once the penalty excess vanishes and the estimate is near $\E_{\min}$,
its $e$ active coordinates have saturated penalties, leaving a vanishing
total penalty for the additional coordinates. Since every
coordinate exceeding a fixed positive multiple of $\delta_n$ pays a cost
bounded away from zero, thresholding removes those additional coordinates
while retaining the minimum-edge support. To omit thresholding, we use
the deletion comparison again, with the nonvanishing ratio
$\lambda_n/\delta_n$ ensuring that the penalty slope dominates the local
likelihood change.

Combining part~\textup{(ii)} with Theorem~\ref{thm:ident}, we obtain
\emph{class selection}, the selection of a graph in the
generating class. If Assumption~\ref{ass:generating}
also holds and $\widehat G_n$ denotes the graph obtained by thresholding an $\eta_n$-approximate global minimizer at $\tau\delta_n$ for a fixed $\tau>0$, then
\[
 \operatorname{pr}\{e(\widehat G_n)=e^*,\ \V_{\widehat G_n}=\V_{G^*}\}\to1.
\]
By Lemma~\ref{lem:ci}, $\widehat G_n$ is therefore Markov equivalent
to $G^*$ with probability tending to one. This class selection requires
neither interiority nor regularity, and by
Theorem~\ref{thm:transfer}\textup{(i)} it extends to every penalty of
Corollary~\ref{cor:family} and every covariance estimate with
$\|S_n-\Sigma^*\|_F=O_p(r_n)$ and $r_n^2=o(\lambda_n)$. Section~S9 of
the Supplementary Material proves this class-selection statement,
including its extension.

A concrete schedule is $\lambda_n=\delta_n=\log(n)/n$ for $n\ge2$,
which places the plateau height at BIC order for the likelihood normalized
per observation and satisfies the ratio condition of
Theorem~\ref{thm:shrinking}\textup{(iii)}. All these
conclusions are for fixed $p$ and a fixed generating parameter, and the
sample size at which they take effect depends on $\Sigma^*$.

\subsection{Covariance estimates and constrained refitting}
\label{sec:refit}
After a support has been selected, we estimate the
parameters on it by maximizing the likelihood again with all other edges
set to zero. Refitting after selection is common for the lasso, where it
removes the shrinkage that the penalty imposes on the retained
coefficients \citep{meinshausen2007,belloni2013}. Our saturated penalty
does not shrink large edges, so here the refit serves a different
purpose. It turns an approximate, thresholded minimizer into a likelihood fit on the selected support; the accuracy of this fit depends only on the covariance estimate and on how precisely the refit is computed. Formally, for a support $H$,
the refit is the problem
\begin{equation}\label{eq:refit}
 \text{minimize } \ell_{S_n}(\theta)\quad\text{over }\theta\in\K
 \text{ with }\supp(\theta)\subseteq H,
\end{equation}
and for a tolerance $\zeta_n\ge0$, a refit with tolerance $\zeta_n$ is any
feasible $\theta$ whose objective exceeds the infimum of
\eqref{eq:refit} by at most $\zeta_n$. The constraint $\theta\in\K$ in \eqref{eq:refit} includes the
eigenvalue constraints $mI\preceq\Omega\preceq M_OI$, and they matter
here. Deleting directed edges keeps $B$ acyclic, whereas setting off-diagonal entries of an error covariance to zero can destroy positive definiteness; optimizing over $\K$ under the zero constraints keeps the refitted error covariance positive definite and well conditioned.

The arguments of Section~\ref{sec:sampling} use the data only through
the sample covariance $S_n$. They therefore apply to any covariance
estimate in place of $S_n$, for instance one computed from non-Gaussian
data (Remark~\ref{rem:fourth_moments}), as long as its error
$\|S_n-\Sigma^*\|_F$ is small.

The next theorem combines the two ideas of this subsection. It states
how small the error of a covariance estimate must be for the selection
results of Section~\ref{sec:sampling} to hold, and it bounds the error
of a refit with tolerance $\zeta_n$ on the selected support.

\begin{theorem}[Support recovery and refitting with estimated covariances]\label{thm:transfer}
Let the penalty be
$P_\delta(\theta)=\sum_k\rho(|a_k(\theta)|/\delta)$ for a function $\rho$
of the saturated family in Corollary~\ref{cor:family}. Suppose a symmetric covariance
estimate satisfies $\|S_n-\Sigma^*\|_F=O_p(r_n)$, $r_n\to0$, for fixed $p$.
Suppose $\lambda_n,\delta_n\to0$, $r_n^2=o(\lambda_n)$ and
$\eta_n=o_p(\lambda_n)$.

\textup{(i)} All $\eta_n$-approximate global minima converge in distance
to $\E_{\min}$ in probability. For each fixed $\tau>0$, thresholding at $\tau\delta_n$
recovers the support of a member of $\E_{\min}$ with probability tending
to one.

\textup{(ii)} Suppose also that every member $(B,\Omega)$ of
$\E_{\min}$ satisfies $mI\prec\Omega\prec M_OI$ and, for some $c_1>0$, $\tau_0\in(0,1)$,
\[
\rho(v)-\rho(u)\ge c_1(v-u),\qquad 0\le u<v\le\tau_0.
\]
If $\liminf\lambda_n/\delta_n>0$, every exact global minimum has the same
support property without thresholding.

\textup{(iii)} Under the generic conditions of Theorem~\ref{thm:ident} and
Corollary~\ref{cor:regular}, a refit on the
support recovered in part~\textup{(i)}, with global objective tolerance
$\zeta_n\to_p0$, has distance $O_p(r_n+\sqrt{\zeta_n})$ to $\E_{\min}$.
\end{theorem}

The condition $r_n^2=o(\lambda_n)$ comes from the comparison
before Theorem~\ref{thm:shrinking}. A covariance error of order $r_n$
changes the best attainable likelihood by order $r_n^2$, and this must be
small compared with the price $\lambda_n$ of one edge. In the refit, the
covariance error enters the local score linearly and the tolerance
enters through its square root, which gives the rate
$r_n+\sqrt{\zeta_n}$. Section~S10 of the Supplementary Material
gives the proof.

By Theorem~\ref{thm:transfer} we may choose the numerical
accuracy of the refit on the same scale as the statistical error, since
$r_n=n^{-1/2}$ and $\zeta_n=O_p(n^{-1})$ preserve the parametric rate.
We measure the tolerance above the infimum over the selected
support, so the rate applies to a local refit only when it lies in
the regular neighborhood and its objective exceeds that of the nearby
minimum-edge representation by at most $\zeta_n$.

\begin{remark}[When the penalty produces exact zeros]\label{rem:raw_support}
We use the slope and ratio conditions in Theorem~\ref{thm:transfer}\textup{(ii)} only for recovery
of support without thresholding. The slope condition asks
$\rho$ to grow at least linearly near zero, as normalized MCP, SCAD and
truncated $\ell_1$ do. The capped power $\rho(s)=\min(s^r,1)$ with $r>1$
does not, since $\rho'(0)=0$. At a representation with $B=0$ and
diagonal $\Omega$, the derivative of $\ell_S$ with respect to an
off-diagonal entry $\Omega_{ij}$ is $-S_{ij}/(\Omega_{ii}\Omega_{jj})$,
which is nonzero whenever the sample cross-covariance $S_{ij}$ is.
Because the penalty has zero slope at $\Omega_{ij}=0$, moving $\Omega_{ij}$ slightly away from zero in the sign of $S_{ij}$ lowers the objective, so this penalty need not
set the entry exactly to zero. Since it still belongs
to the saturated family, however, the thresholded conclusion of part~(i) applies.
This parallels the theory of nonconvex penalties for regression,
where \citet{lohwainwright2015} bound the estimation error of every
stationary point and \citet{loh2017} obtain support recovery for penalties
whose derivative at the origin is positive. In our setting the
deletion argument must also preserve an acyclic coefficient matrix and a
positive definite error covariance.
\end{remark}

\begin{remark}[Sampling beyond the Gaussian model]\label{rem:fourth_moments}
For independent and identically distributed observations with finite fourth
moments, the centered empirical covariance has error $O_p(n^{-1/2})$, so
Theorem~\ref{thm:transfer} applies with $r_n=n^{-1/2}$. The recovered
object is then the equivalence class that the Gaussian model assigns to
the population covariance, which need not describe every conditional
independence of the non-Gaussian sampling distribution.
\end{remark}

\section{\texorpdfstring{Separation, signal strength and the choice of penalty}{Separation, signal strength and the choice of penalty}}\label{sec:separation}
Section~\ref{sec:estimation} shows that, as $n$ grows, the penalized likelihood achieves class selection (Section~\ref{sec:sampling}), selecting a graph in the generating algebraic class of $G^*$. Those results are
asymptotic and do not say how large $n$ must be. The answer depends on
how well other graphs can imitate $\Sigma^*$. By Theorem~\ref{thm:ident}, no
graph outside the generating class with at most $e^*$ edges represents
$\Sigma^*$ exactly, but the covariance model of such a graph can come very
close to $\Sigma^*$ when $\theta^*$ is near an exceptional set, and then many
observations are needed to tell the two apart. In
Section~\ref{sec:separation_setting} we measure how far these competing
models stay from $\Sigma^*$ and give a condition under which an
approximate minimizer selects the generating class. In
Section~\ref{sec:lowerbound} we show that no estimator can select the
class when the signal is too weak, and in Section~\ref{sec:l1} we compare
the saturated and $\ell_1$ penalties.

\subsection{A quantitative separation condition}\label{sec:separation_setting}
The argument compares two distances. If the selected graph has at most $e$ edges and lies outside
the generating class, every covariance it produces is at least a fixed
distance $\Delta$ from $\Sigma^*$, which we call the \emph{separation}. On
the other hand, the covariance fitted by an approximate minimizer is close
to $\Sigma^*$ when $S$ is close to $\Sigma^*$ and the tuning parameters are
small. When the second distance is smaller than $\Delta$, the selected
graph cannot lie outside the class. The setting below makes both
distances explicit. As in Theorem~\ref{thm:ident}, the generating graph
is bow-free and its parameter is generic, so $e=e^*$; the competing
graphs may contain bows.

\noindent\textit{Setting for Theorem~\ref{thm:certificate}.}\label{setting:separation}
Impose Assumptions~\ref{ass:generating} and~\ref{ass:domain}, let $[a,b]$, with $a>0$, contain
the eigenvalues of every covariance in $\Phi(\K)$, and let $S\succ0$ have
eigenvalues in $[a/2,2b]$. Let $\widehat\theta$ be an $\eta$-approximate global
minimizer of the quasi-MCP objective $\ell_S+\lambda P_\delta$, and define
$A(S)=\{\operatorname{tr}(S\Sigma^{*-1})-\log\det(S\Sigma^{*-1})-p\}/2$,
$r_p=\sqrt{p(p-1)}$, and $c_p=\sqrt{2p(p-1)}$, together with
\[
u=\frac{A(S)+\eta}{\lambda},\qquad
R=\|S-\Sigma^*\|_F+\frac{16b^2}{a}\sqrt{A(S)+\lambda e+\eta}.
\]
Delete edge coordinates smaller than $\delta$ in magnitude while retaining
the error variances, and denote the resulting parameter by $\bar\theta$
and its support by $\widehat H$. The enlarged domain $\K^+$ has the
original coefficient-matrix domain and the relaxed eigenvalue
constraints $mI/2\preceq\Omega\preceq(M_O+m/2)I$; when $r_p\delta\le m/2$,
the deletion stays inside $\K^+$. Let
$L_\Phi$ be a Lipschitz constant of $\Phi$ along the segment from
$\widehat\theta$ to $\bar\theta$; explicit values of $a$, $b$ and $L_\Phi$ that depend only on $p$,
$M_B$, $M_O$ and $m$ are given in Section S11 of the Supplementary
Material \citep{supplement}, together with the proofs for this subsection. Define
the separation by
\[
\Delta=\min_{\substack{H:e(H)\le e\\\V_H\ne\V_{G^*}}}
\inf_{\substack{\theta\in\K^+\\\supp(\theta)\subseteq H}}
 \|\Phi(\theta)-\Sigma^*\|_F,
\]
with $\Delta=\infty$ when the outer collection is empty; generic
identification and compactness of $\K^+$ imply $\Delta>0$.

The remaining quantities bound how far the fitted covariance
can be from $\Sigma^*$. The term $A(S)$ is the largest likelihood gain
that the sampling error in $S$ makes available; it is zero when
$S=\Sigma^*$ and of order $n^{-1}$ for a sample covariance. The ratio $u$
compares this gain, plus the optimization error $\eta$, with the price
$\lambda$ of one edge. The radius $R$ bounds the distance between
$\Sigma^*$ and the covariance of $\widehat\theta$, and the eigenvalue range
$[a,b]$ of the covariances in $\Phi(\K)$ enters it through the curvature
of the likelihood. Thresholding then moves the covariance by at most $L_\Phi c_p\delta$, where $r_p$ and $c_p$ are the Frobenius-norm factors for the deleted off-diagonal entries of $\Omega$ and of $(B,\Omega)$.

Theorem~\ref{thm:certificate} combines these quantities in three
conditions. The condition $u<1$ ensures that at most $e$ edges survive
thresholding, because each additional edge costs a full $\lambda$. The
condition $r_p\delta\le m/2$ keeps the thresholded error covariance within
the relaxed eigenvalue constraints of $\K^+$. The main condition,
$R+L_\Phi c_p\delta<\Delta$, says that the covariance of the thresholded
estimate is closer to $\Sigma^*$ than any competing class can come, so its
graph lies in the generating class, and the dimension count of
Theorem~\ref{thm:ident} then forces exactly $e$ edges.

\begin{theorem}[Selection under covariance-model separation]\label{thm:certificate}
In the setting of Section~\ref{sec:separation_setting}, if $u<1$,
$r_p\delta\le m/2$, and
\[
R+L_\Phi c_p\delta<\Delta,
\]
then every thresholded approximate minimum defined there has exactly $e$
edges and belongs to the generating algebraic class, and hence to the
generating Markov class. Its error covariance is positive definite, and
the thresholded estimate lies in $\K^+$.
\end{theorem}

The condition of Theorem~\ref{thm:certificate} involves the
population quantities $e$ and $\Delta$, so it cannot be checked from data,
and the explicit constant $L_\Phi$ can be conservative when $pM_B$ is
large. We can still derive from the condition an explicit range of
tuning parameters for class selection. Suppose that $e\ge1$ and
$\Delta<\infty$, and put
\[
\bar\lambda=\frac1{16e}\Big(\frac{a\Delta}{16b^2}\Big)^2,\qquad
\bar\delta=\min\Big\{\frac{\Delta}{4L_\Phi c_p},\frac{m}{2r_p}\Big\}.
\]
Let $S_n$ be the Gaussian sample covariance with $p$ fixed, as in
Assumption~\ref{ass:tuning}. If $\eta_n\to_p0$, then for each fixed
$\lambda\in(0,\bar\lambda]$ and $\delta\in(0,\bar\delta]$ the hypotheses
of Theorem~\ref{thm:certificate} hold with probability tending to one;
if $\eta_n=o_p(\lambda_n)$, the same is true for the schedule
$\lambda_n=\delta_n=\log(n)/n$. In both cases the thresholded approximate
global minimizer has exactly $e$ edges and lies in the generating
algebraic class with probability tending to one.

\begin{remark}[Class selection and refitting]\label{rem:class_refit}
Theorem~\ref{thm:certificate} identifies only an algebraic
class. It does not show that the selected support carries an exact
positive definite representation of $\Sigma^*$. That stronger conclusion,
recovery of a support in $\E_{\min}$, and the refit rate follow under
the conditions of Theorem~\ref{thm:transfer}. The two results depend on
different quantities. The separation $\Delta$ measures how close
competing classes come to $\Sigma^*$, while the smallest eigenvalue of the
information matrix restricted to a recovered regular support controls
the accuracy of the parameters on that support.
\end{remark}

\subsection{Weak correlations and class-selection error}\label{sec:lowerbound}
Section~\ref{sec:separation_setting} gives a condition under
which the estimator selects the generating class. We now show that a
condition of this kind is necessary. When the correlations are too weak
for the sample size, no estimator can select the class reliably. We use
the simplest family in which only the class has to be found. Among $p$
variables, exactly one pair $k=(i,j)$ is correlated, with correlation $r$,
and all other variables are independent. This is the covariance of
Example~\ref{ex:two_variable} placed among $p$ variables. Its sparsest
representations have a single edge between $i$ and $j$, which may be
$i\to j$, $j\to i$ or $i\leftrightarrow j$, and these three graphs are
algebraically equivalent, so the class $\mathcal C_k$ is determined by the
pair $k$ and not by the type of edge. Selecting the class therefore means
finding the correlated pair among $M=\binom p2$ candidates, a testing
problem among $M$ hypotheses that does not ask for directions the
covariance cannot identify. In the next proposition, $e_1,\dots,e_p$ are the standard basis
vectors of $\mathbb R^p$, and
$\operatorname{pr}_k$ denotes probability when the observations are drawn
from the $k$th model $N(0,\Sigma_k)$.

\begin{proposition}[Correlated-pair testing bound]\label{prop:fano}
For $p\ge3$, let $M=\binom p2$ and index unordered pairs by $k=(i,j)$.
The Gaussian covariances
\[
\Sigma_k=I+r(e_ie_j^T+e_je_i^T),\qquad 0<r\le1/2,
\]
have minimum joint edge count one and distinct algebraic/Markov classes
$\mathcal C_k$. For every class estimator based on $n$ observations,
\[
\sup_k \operatorname{pr}_k(\widehat{\mathcal C}\ne\mathcal C_k)
\ge1-\frac{nr^2+\log2}{\log M}.
\]
For the two-variable comparison of independence with $\Sigma_r$, the maximum
classification error is at least $3/8$ if $nr^2\le1/8$.
\end{proposition}

The bound is a standard application of Fano's inequality.
Since $n$ observations carry a Kullback--Leibler divergence of at most
$nr^2$ about which pair is correlated, they cannot distinguish the $M$
pairs unless $nr^2$ is at least of order $\log M$ (Section~S12 of the
Supplementary Material).

The condition of Section~\ref{sec:separation_setting} can be computed for
the same family, which lets us compare the two results. The competitors
of the pair $k=(i,j)$ are the empty graph and the single edges on other
pairs. Their covariances have a zero $(i,j)$ entry, so the separation is
$\Delta=\sqrt2\,r$, attained at the identity, and $e=1$. These explicit
values replace the genericity assumption. Fix $p$ and a domain with
$m\le1/2$ and $M_O\ge3/2$, which contains every $(0,\Sigma_k)$, and take
$\lambda_n=\delta_n=\omega_n/n$, where $\omega_n\to\infty$,
$\omega_n/n\to0$ and $\eta_n=o(\lambda_n)$. The distribution of $A(S_n)$
does not depend on $k$, and $A(S_n)=O_p(n^{-1})$, so
Theorem~\ref{thm:certificate} applies to the whole family at once. If the
correlation is at least $\underline r_n$ with
$n\underline r_n^2/\omega_n\to\infty$, the thresholded approximate
minimizer has exactly one edge, joining $i$ and $j$, with probability
tending to one, uniformly in $k$ and in $r\in[\underline r_n,1/2]$.

Together, the two results locate the smallest correlation at which the
class can be selected. For fixed $p$, every estimator fails when $nr^2$ is
below a constant times $\log M$, by Proposition~\ref{prop:fano}, and the
penalized likelihood succeeds when $nr^2$ is larger than $\omega_n$. The
critical correlation is therefore of order $n^{-1/2}$, up to the factor
$\omega_n^{1/2}$, which can grow arbitrarily slowly and equals
$(\log n)^{1/2}$ for the schedule $\lambda_n=\delta_n=\log(n)/n$. The two
bounds do not match exactly, so they do not give a minimax rate
(Section~S12 of the Supplementary Material).

\subsection{Comparison with \texorpdfstring{$\ell_1$}{l1} regularization}\label{sec:l1}
An $\ell_1$ penalty differs from a saturated one in two ways. It
shrinks parameters within a regular support, and along a fiber it can
favor representations with a different joint edge count. For directed acyclic graphs, \citet{jin2026} show
that an $\ell_1$-penalized Gaussian likelihood can miss the Markov
equivalence class even at its global optimum, and replace it by an
$\ell_0$ penalty. Proposition~\ref{prop:l1} and
Example~\ref{ex:collider} below show that the same failure occurs for
mixed graphs, where the penalty acts on coefficients and error
covariances jointly. Section~S13 of the Supplementary Material gives
the calculations for this subsection.

For the first comparison, recall that $a(\theta)$ lists the edge coordinates, let $R_1(\theta)=\sum_k|a_k(\theta)|$ be their joint $\ell_1$ norm, and let $\kappa>0$ be the weight of
this penalty in the objective $\ell+\kappa R_1$, the analog of $\lambda$.
An interior exact representation on a regular support with an active
edge is not stationary for $\ell+\kappa R_1$, since shrinking that edge
lowers the penalty at first order and changes the likelihood only at
second order. The nearby stationary point with the same edge signs is
displaced by $O(\kappa)$, so a weight
$\kappa_n$ adds a bias of order $\kappa_n$ to the estimate.
\citet{deng} give a bivariate directed acyclic example of this
shrinkage, which a saturated penalty avoids.

For the second comparison, we determine which exact representations the
estimator approaches when the $\ell_1$ penalty vanishes slowly enough to
dominate the available likelihood improvement.

\begin{proposition}[Minimum-$\ell_1$ representations]\label{prop:l1}
If $\kappa_n\to0$, $S_n\to_p\Sigma^*$ and
$(A(S_n)+\eta_n)/\kappa_n\to_p0$, approximate global minima of
$\ell_{S_n}+\kappa_nR_1$ approach
$\E_1=\argmin_{\theta\in\E}R_1(\theta)$. In general
$\E_1\not\subseteq\E_{\min}$ on the broad acyclic mixed domain.
\end{proposition}

\begin{figure}[t]
\centering
\includegraphics[width=\linewidth]{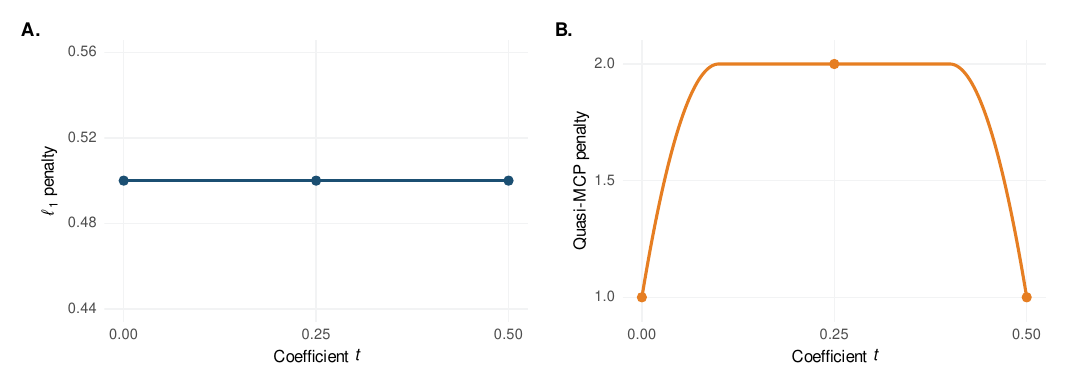}
\caption{\textbf{Penalty Values Along a Covariance Fiber.}
The path in Example~\ref{ex:two_variable} preserves the standardized
covariance with correlation $r=1/2$.
\textbf{A.} The joint $\ell_1$ penalty equals $1/2$ throughout the path,
so the two-edge bowed point at $t=1/4$ (middle dot) ties with the
one-edge endpoints (outer dots).
\textbf{B.} Quasi-MCP with $\delta=0.1$ has penalty one at the endpoints
and two on $[0.1,0.4]$, whose interior points are suboptimal local minima
of the population objective.}
\label{fig:fiber}
\end{figure}

In the two-variable family $(B_t,\Omega_t)$ of
Example~\ref{ex:two_variable}, with bounds that contain the path
for $0\le t\le r$, every point of the path has joint $\ell_1$ value $|t|+|r-t|=r$, so
the one-edge endpoints tie with the bowed interior points
(Figure~\ref{fig:fiber}\textbf{A}) and the $\ell_1$ target can contain
a redundant edge. All these representations lie in one algebraic
class (Section~\ref{sec:equivalence}).

With three variables the $\ell_1$ target can leave the generating
class.

\begin{example}[An $\ell_1$ target outside the generating class]\label{ex:collider}
Let $G^*$ be the collider $1\to3\leftarrow2$ with
$B^*_{13}=B^*_{23}=\gamma>0$ and $\Omega^*=I$, so that
$\sigma^*_{12}=0$,
$\sigma^*_{13}=\sigma^*_{23}=\gamma$ and $\sigma^*_{33}=1+2\gamma^2$. A graph in
the algebraic class of $G^*$ is Markov equivalent to $G^*$
(Lemma~\ref{lem:ci}), so it has no edge on $\{1,2\}$ and every edge on
$\{1,3\}$ or $\{2,3\}$ has an arrowhead at $3$. Matching
$\sigma^*_{13}$ and $\sigma^*_{23}$, every exact representation on such a
graph has $B_{13}=t_1$, $B_{23}=t_2$, $\Omega_{13}=\gamma-t_1$,
$\Omega_{23}=\gamma-t_2$ and $\Omega_{12}=0$ for real $t_1,t_2$. Hence
\[
R_1=|t_1|+|\gamma-t_1|+|t_2|+|\gamma-t_2|\ge2\gamma,
\]
with equality if and only if $t_1,t_2\in[0,\gamma]$. In the unrestricted
class $\mathcal A_0$, the minimum-edge fiber
consists of the four corners $t_1,t_2\in\{0,\gamma\}$, with graphs
$1\to3\leftarrow2$, $1\leftrightarrow3\leftarrow2$,
$1\to3\leftrightarrow2$ and $1\leftrightarrow3\leftrightarrow2$. Now let
$\tilde\theta$ have $\tilde B_{31}=\tilde B_{32}=\gamma/(1+2\gamma^2)$,
$\tilde\Omega_{11}=\tilde\Omega_{22}=(1+\gamma^2)/(1+2\gamma^2)$,
$\tilde\Omega_{12}=-\gamma^2/(1+2\gamma^2)$, $\tilde\Omega_{33}=1+2\gamma^2$ and
$\tilde\Omega_{13}=\tilde\Omega_{23}=0$. It is an exact bow-free
representation with the complete graph $3\to1$, $3\to2$,
$1\leftrightarrow2$, and
\[
R_1(\tilde\theta)=\frac{\gamma(2+\gamma)}{1+2\gamma^2}
=2\gamma-\frac{\gamma^2(4\gamma-1)}{1+2\gamma^2}.
\]
Hence, if $\gamma>1/4$ and the bounds of Assumption~\ref{ass:domain}
contain $\tilde\theta$, no member of $\E_1$ has its graph in the algebraic
or Markov class of $G^*$, and in particular $\E_1\cap\E_{\min}=\emptyset$;
at $\gamma=1/4$ the two values tie. For $\gamma=1/2$ we have
$R_1(\tilde\theta)=5/6<1=2\gamma$, and the bounds $M_B=1$, $m=1/4$ and
$M_O=2$ contain $\tilde\theta$ and all four minimum-edge representations.
\end{example}

When $\gamma>1/4$, $\E_1$ lies at a positive distance from the
compact set of exact representations in $\K$ with graphs in the
generating class, so under the conditions of Proposition~\ref{prop:l1}
the approximate global minima of $\ell_{S_n}+\kappa_nR_1$ stay bounded
away from this set with probability tending to one.
Since $\tilde\theta$ is bow-free, the same holds on a bow-free domain. A
saturated penalty selects $\E_{\min}$
here (Theorem~\ref{thm:penalty}). The example depends on the units of
the variables, because $R_1$ adds coefficients and error covariances, which change
differently under rescaling, and once $X_3$ is standardized the
representation $\tilde\theta$ has a larger $R_1$ than the class
representations. The joint edge count does
not depend on the units, and rescaling does not change the graphs in the
unrestricted minimum-edge fiber. Standardizing does not remove the
failure, however, although on three standardized variables a
numerical search over all algebraic classes whose variety is not the
whole space of symmetric matrices found no such example. On four
variables it occurs. Let $G^*$ be the diamond $1\to2$, $1\to3$, $2\to4$,
$3\to4$ with $B^*_{12}=B^*_{13}=1/2$, $B^*_{24}=4/5$, $B^*_{34}=-3/5$ and
error variances $1$, $3/4$, $3/4$ and $6/25$, so that every variable has
unit variance. Every exact representation whose graph is Markov
equivalent to $G^*$ has $R_1\ge12/5$, the value at $\theta^*$, while the
exact bow-free representation with $B_{21}=B_{31}=2/5$ on the graph
$2\to1$, $3\to1$, $2\leftrightarrow3$, $2\leftrightarrow4$,
$3\leftrightarrow4$ has $R_1=21/10$ (Section S13 of the Supplementary
Material).

On the two-variable path we also see that an $\ell_1$ solution need not
provide an initial point from which local descent on the saturated criterion reaches a minimum-edge representation. At an interior point with
$\delta<\min(t,r-t)$, both active penalties stay saturated nearby and
acyclicity excludes the reverse directed edge; since this exact
representation minimizes the population likelihood, it is a local
minimum of the saturated objective, with value $\ell_*+2\lambda$, one
$\lambda$ more than either endpoint.
Figure~\ref{fig:fiber}\textbf{B} displays these score levels.
Saturation separates the redundant representation from the endpoints
globally and leaves an interior local minimum that a local algorithm
must escape.
\section{Computation by alternating minimization}\label{sec:computation}
\begin{figure}[t]
\centering
\includegraphics[width=\linewidth]{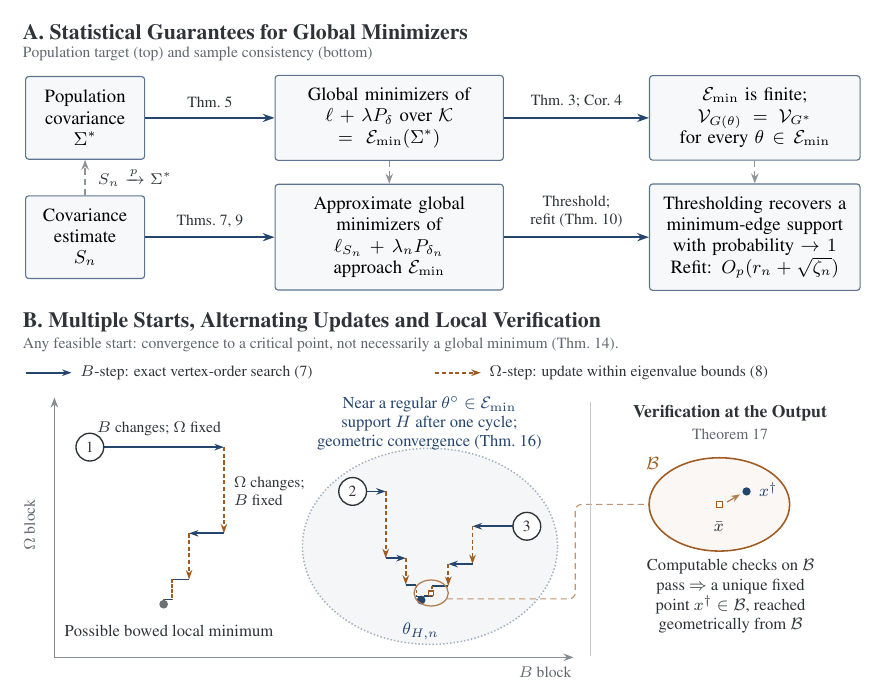}
\caption{\textbf{Guarantees for the Global Minimizer and for the Alternating Algorithm.}
\textbf{A.} For a generic parameter of a bow-free generating graph, or
more generally when $\dim\V_{G^*}=p+e^*$, the global minimizers of the
penalized population objective are the sparsest exact representations,
and their graphs lie in the generating algebraic class, even though
competing graphs may contain bows. With a covariance estimate $S_n$,
approximate global minimizers approach them, thresholding recovers a
minimum-edge support, and refitting attains the rate $O_p(r_n+\sqrt{\zeta_n})$ of Theorem~\ref{thm:transfer}. A joint
$\ell_1$ penalty, by contrast, can select a graph outside the generating
Markov class (Proposition~\ref{prop:l1} and Example~\ref{ex:collider}).
\textbf{B.} A schematic view of the alternating algorithm
in the space of the two parameter blocks. Circled numbers mark feasible
starts. Each $B$-step moves horizontally and each $\Omega$-step
vertically. Start 1 reaches a critical point that may be a bowed local
minimum. Starts 2 and 3 lie in the shaded neighborhood of a regular
sparsest representation, where the algorithm finds its support after one
cycle and converges geometrically to the sample solution $\theta_{H,n}$.
The inset shows the certificate around a computed output $\bar x$
(square).}
\label{fig:guarantees}
\end{figure}

Sections~\ref{sec:estimation} and~\ref{sec:separation}
concern global minimizers of $\ell_{S_n}+\lambda P_\delta$, but this
objective is not convex, so finding a global minimizer is a problem of
its own. Figure~\ref{fig:guarantees} contrasts what the
statistical results give for a global minimizer with what the algorithm
guarantees. The objective is not even biconvex in $(B,\Omega)$. For a fixed
error covariance, the likelihood is a convex quadratic function of $B$,
but acyclicity restricts $B$ to a nonconvex set and the saturated
penalty is concave near zero. For a fixed coefficient matrix, the
likelihood is convex in the precision matrix $\Omega^{-1}$ but not in
$\Omega$ itself. We therefore update $B$ and $\Omega$ in alternating
proximal steps, which keep each update tractable. For a fixed mixed
graph, the standard maximum likelihood algorithms update the coefficients
and error covariances attached to one vertex at a time
\citep{ricf2009,dfw2019}. Our blocks are the whole matrices $B$ and
$\Omega$, which suits the penalty. The $B$-step then reduces to a search
over vertex orders, and for small step sizes the $\Omega$-step is a
strongly convex problem. Algorithm~\ref{alg:estimator} summarizes the
procedure.

Methods that also learn the graph search over mixed graphs
greedily, as \citet{nowzohour} do for bow-free graphs, exactly, as
\citet{rantanen2021} do for maximal ancestral graphs, or by continuous
optimization in the tradition of NOTEARS \citep{notears2018}, which
encodes acyclicity and the graph class as smooth constraints
\citep{dcd,ashman2023,spot}. DECOR \citep{decor} belongs to the last
group. It uses the smooth acyclicity constraint of NOTEARS and
encourages bow-freeness only through an optional penalty or a post hoc
reconciliation step. Our algorithm keeps DECOR's alternation between $B$ and $\Omega$, but computes each update of $B$ exactly, by searching over vertex
orders with the subset dynamic program of exact structure search
\citep{koivisto2004,silander2006}, and ends by refitting the likelihood on
the selected graph, as in Section~\ref{sec:refit}. Already for two
responses with correlated errors, each regressed on its own covariate,
the likelihood can have several local maxima
\citep{drtonrichardson2004}, so guarantees for iterative fitting, ours
included, are local.

Recall that the domain $\K$ of Assumption~\ref{ass:domain} permits
bows and is the product of a coefficient domain $\K_B$ and an
error-covariance domain $\K_\Omega$. We split the penalty as
$P_\delta=P_{\delta,B}+P_{\delta,\Omega}$, with
$P_{\delta,B}(B)=\sum_{i\ne j}q_\delta(B_{ij})$ and
$P_{\delta,\Omega}(\Omega)=\sum_{i<j}q_\delta(\Omega_{ij})$, and write
$g_B=\lambda P_{\delta,B}+\mathbf1_{\K_B}^{\infty}$ and
$g_\Omega=\lambda P_{\delta,\Omega}+\mathbf1_{\K_\Omega}^{\infty}$,
where the indicator $\mathbf1_A^\infty$ is zero on $A$ and infinite
elsewhere. For step sizes $\alpha,\beta>0$,
one cycle consists of
\begin{align}
B^{t+1}&\in\argmin_U\left\{
 \langle\nabla_B\ell_{S_n}(B^t,\Omega^t),U-B^t\rangle
 +\frac{\|U-B^t\|_F^2}{2\alpha}+g_B(U)\right\},\label{eq:prox}\\
\Omega^{t+1}&\in\argmin_V\left\{
 \langle\nabla_\Omega\ell_{S_n}(B^{t+1},\Omega^t),V-\Omega^t\rangle
 +\frac{\|V-\Omega^t\|_F^2}{2\beta}+g_\Omega(V)\right\}.
 \label{eq:noiseprox}
\end{align}
Here $\langle A,C\rangle=\operatorname{tr}(A^TC)$, and
$\nabla_\Omega\ell_{S_n}$ is the gradient for this inner product on
symmetric matrices. Both steps preserve feasibility.

Let $L_B,L_\Omega$ be block gradient
Lipschitz constants of $\ell_{S_n}$ on a bounded convex neighborhood of $\K$ with uniformly
positive definite error covariance. In
Theorem~\ref{thm:optimization} below we assume $0<\alpha<1/L_B$ and
$0<\beta<1/L_\Omega$. These constants grow as the lower eigenvalue bound $m$
decreases, so the admissible global steps can be very small; the
local-convergence illustration of Section~\ref{sec:local_numerics}
uses larger steps, which meet the contraction conditions stated after Theorem~\ref{thm:local}.

\begin{algorithm}[t]
\caption{Jointly Sparse Covariance Representation}\label{alg:estimator}
\begin{enumerate}
\item Input a covariance estimate $S_n$, tuning parameters $\lambda,\delta$, step sizes
$\alpha,\beta$, parameter bounds, and a feasible initial pair $(B^0,\Omega^0)$.
\item Repeat the $B$-step \eqref{eq:prox}, minimizing its proximal objective
over vertex orders, followed by the $\Omega$-step \eqref{eq:noiseprox}
evaluated at the newly updated coefficient matrix, until the cycle
length $\|\theta^{t+1}-\theta^t\|_F$ falls below a tolerance.
\item Delete edge coordinates whose magnitude is below $\tau\delta$ for a
fixed $\tau>0$ (with fixed tuning parameters, $\tau\delta$ should lie below the
magnitudes of the edges to be retained).
\item Refit the likelihood over $\K$ with the selected zero constraints,
and return the resulting representation and support graph.
\end{enumerate}
\end{algorithm}

\subsection{Solving the two steps}\label{sec:steps}
We now compute the two steps, beginning with a single
entry of the $B$-step. Suppose first that $\alpha<\delta^2/(2\lambda)$, so that the scalar
proximal objective of the $B$-step is strongly convex, and put
$z_0=2\alpha\lambda/\delta$ and $k=1-2\alpha\lambda/\delta^2$; if $z$ is an
entry of $B-\alpha\nabla_B\ell_{S_n}(B,\Omega)$, the
unconstrained scalar solution is
\[
 t(z)=\operatorname{sign}(z)
 \begin{cases}0,&|z|\le z_0,\\ (|z|-z_0)/k,&z_0<|z|<\delta,\\
 |z|,&|z|\ge\delta.
 \end{cases}
\]
Clipping this solution to $[-M_B,M_B]$ enforces the coefficient bound.

For a fixed vertex order, the $B$-step separates over entries pointing
forward in that order, so we can first compute their scalar gains and then
optimize the order itself, as permutation-based methods for directed
acyclic graphs do \citep{raskutti,solus2021}. Write $[p]=\{1,\dots,p\}$, let $w_{ij}$ denote the reduction in the $(i,j)$ scalar proximal objective from allowing $B_{ij}\in[-M_B,M_B]$ rather than $B_{ij}=0$, and let
$D(A)$ be the largest total gain among orders of a vertex set
$A\subseteq[p]$. If we place a vertex $j$ last in $A$, its incoming
directed edges contribute independently of the order chosen for the
remaining vertices, which gives the subset recursion used in exact
structure learning of directed acyclic graphs \citep{koivisto2004,silander2006}:
\[
 D(A)=\max_{j\in A}\left\{D(A\setminus\{j\})+
                  \sum_{i\in A\setminus\{j\}}w_{ij}\right\},
 \qquad D(\varnothing)=0.
\]
Because every acyclic graph admits a topological order, backtracking
from $[p]$ through the maximizing last vertices gives an order whose
clipped scalar solutions form an exact $B$-step. The recursion costs $O(p^22^p)$
operations and $O(2^p)$ storage (Algorithm~S1 in Section~S14 of the
Supplementary Material, \citealp{supplement}). Maximizing a sum of pairwise gains over vertex orders is a
linear ordering problem, NP-hard for general weights, so an exponential
worst-case cost is expected; since the recursion runs in every cycle, the
exact $B$-step is practical for at most a few tens of variables.

For the $\Omega$-step, with $\nabla_\Omega$ the gradient for the
trace inner product on symmetric matrices, the Frobenius norm counts each
off-diagonal pair twice and the penalty counts it once, so a pair behaves
like a scalar update with step $\beta/2$. For $\beta<\delta^2/\lambda$ the
$\Omega$-step is therefore strongly convex, and its entrywise solution zeroes inputs of
magnitude at most $\beta\lambda/\delta$ and leaves those of magnitude at least $\delta$
unchanged. The eigenvalue
constraints couple the entries, so this solution is the exact update
precisely when it satisfies $mI\preceq V\preceq M_OI$; otherwise we
compute the update with a conic solver and rely on
Proposition~\ref{prop:inexact}.

These closed forms need $\alpha<\delta^2/(2\lambda)$ and
$\beta<\delta^2/\lambda$, which under the schedule
$\lambda_n=\delta_n=\log(n)/n$ of Section~\ref{sec:sampling} read
$\alpha<\log(n)/(2n)$ and $\beta<\log(n)/n$. The $B$-step does not need
this restriction. For $\alpha\ge\delta^2/(2\lambda)$ the scalar minimizer
is the hard threshold at $(2\alpha\lambda)^{1/2}$, whose clipping to
$[-M_B,M_B]$ is exact when $M_B\ge(2\alpha\lambda)^{1/2}$, as holds for
large $n$ under the schedule. With scalar gains computed under $|B_{ij}|\le M_B$, the
recursion returns an exact $B$-step for every $\alpha$. Section~S14 of the
Supplementary Material
also derives the closed forms of this subsection.

\subsection{Convergence and local support recovery}\label{sec:convergence}
We apply the proximal alternating linearized minimization framework of
\citet{palm} and the descent theory of \citet{attouch2010} and
\citet{abs}.
For an objective $J$, we call $x$ a \emph{limiting critical point}
when $0\in\partial J(x)$ for the limiting subdifferential $\partial$, and
we say that a sequence $(\theta^t)$ has \emph{finite length} when
$\sum_t\|\theta^{t+1}-\theta^t\|_F<\infty$.

\begin{theorem}[Alternating descent and stationarity]\label{thm:optimization}
Every exact sequence of \eqref{eq:prox}--\eqref{eq:noiseprox} converges to a
limiting critical point of $F_n+\mathbf1_\K^\infty$ and has finite length.
With $a_B=(\alpha^{-1}-L_B)/2>0$ and
$a_\Omega=(\beta^{-1}-L_\Omega)/2>0$,
\[
F_n(\theta^t)-F_n(\theta^{t+1})\ge
 a_B\|B^{t+1}-B^t\|_F^2+a_\Omega\|\Omega^{t+1}-\Omega^t\|_F^2.
\]
Consequently the smallest squared cycle length through $T$ cycles is bounded
by $(F_n(\theta^0)-\min_\K F_n)/(T\min(a_B,a_\Omega))$.
\end{theorem}

Section~S14 of the Supplementary Material gives the proof,
which verifies the Kurdyka--\L{}ojasiewicz property of the objective on
the union of acyclic supports.

Computed block updates are approximate, and we measure their
accuracy relative to the cycle length. Write $\Delta B=B^{t+1}-B^t$ and
$\Delta\Omega=\Omega^{t+1}-\Omega^t$.

\begin{proposition}[Convergence with inexact block updates]\label{prop:inexact}
The stationarity conclusion remains valid if each feasible block output has
proximal objective no larger than the old block, and there exist
$\xi_B\in\partial g_B(B^{t+1})$,
$\xi_\Omega\in\partial g_\Omega(\Omega^{t+1})$ such that
\begin{align*}
&\|\nabla_B\ell_{S_n}(B^t,\Omega^t)+\Delta B/\alpha+\xi_B\|_F\\
&\quad+\|\nabla_\Omega\ell_{S_n}(B^{t+1},\Omega^t)
                      +\Delta\Omega/\beta+\xi_\Omega\|_F
\le c\|\theta^{t+1}-\theta^t\|_F
\end{align*}
for a fixed finite $c$.
\end{proposition}

The residual must be small relative to the cycle length; a
solver's absolute stopping tolerance does not by itself ensure this
(Section~S15 of the Supplementary Material).

Theorem~\ref{thm:optimization} and Proposition~\ref{prop:inexact}
guarantee convergence to a critical point from any feasible start, but
not to a minimum-edge representation. A stronger conclusion holds near a
regular minimum-edge representation, where quasi-MCP removes inactive
entries while leaving the active likelihood equations unchanged.

\begin{theorem}[Local support recovery and geometric convergence]\label{thm:local}
Let $\theta^\circ\in\E_{\min}$ be regular and interior as in
Corollary~\ref{cor:regular}, with support $H$. Fix quasi-MCP tuning parameters as in
Theorem~\ref{thm:penalty} with $\delta$ below every nonzero edge magnitude
of $\theta^\circ$. Let $\theta_{H,n}$ minimize $\ell_{S_n}$ over parameters
supported in $H$ near $\theta^\circ$, and let $I_H$ be the restricted
information, the Hessian of $\ell$ at $\theta^\circ$ in orthonormal free
coordinates of $H$ for the product Frobenius norm, with
$\|h\|_{I_H}=(h^TI_Hh)^{1/2}$. For sufficiently small fixed $\alpha,\beta$,
with probability tending to one there is a neighborhood of
$\theta_{H,n}$ such that every exact
alternating sequence initialized there has the following properties.
\begin{enumerate}
\item[(i)] The sequence identifies the support $H$ after
one cycle.
\item[(ii)] The sequence converges geometrically. For some $q<1$ and all $t\ge1$,
\[
\|\theta^t-\theta^\circ\|_{I_H}
\le q^{t-1}\|\theta^1-\theta_{H,n}\|_{I_H}+O_p(n^{-1/2}).
\]
Norm equivalence gives the same conclusion in Frobenius norm with fixed
multiplicative constants.
\end{enumerate}
\end{theorem}

The two parts of the theorem correspond to two stages of the
proof. In the first cycle, the scalar thresholds of the two steps and the
comparison of vertex orders remove the small inactive entries, which
fixes the support. After that the penalty is constant, and the cycle is
a smooth map that contracts in the information norm $\|\cdot\|_{I_H}$.
The step-size condition can be made explicit. With $A$ and $D$ the
diagonal blocks of $I_H$ for $B$ and for $\Omega$, the contraction holds with
probability tending to one whenever $\alpha\lambda_{\max}(A)<2$ and
$\beta\lambda_{\max}(D)<2$ (Section~S16 of the Supplementary Material);
a small $\alpha$ is still needed for the comparison of orders. When the
tuning parameters shrink with $n$, keeping the $\Omega$-step strongly
convex requires $\beta<\delta_n^2/\lambda_n$ (Section~\ref{sec:steps}), so
under the schedule $\lambda_n=\delta_n=\log(n)/n$ the step $\beta$ must
shrink as well, and $q$ need not be uniform in $n$.

\begin{remark}[Initialization and basin entry]\label{rem:basin}
The starting point in Theorem~\ref{thm:local} may carry small
additional directed or bidirected edges, provided it lies in the stated
neighborhood.
If that neighborhood contains the Frobenius ball of radius $R_n$
around $\theta_{H,n}$, then by the triangle inequality a feasible
initializer $W_n$ enters it whenever
\[
\|W_n-\theta^\circ\|_F+\|\theta_{H,n}-\theta^\circ\|_F<R_n.
\]
By Proposition~\ref{prop:l1} and Example~\ref{ex:collider}, preliminary
$\ell_1$ minimizers can stay at a positive
distance from $\E_{\min}$, so saturated refinement of an $\ell_1$
initializer, followed by a refit meeting the objective tolerance of
Theorem~\ref{thm:transfer}, is justified once the initializer satisfies
this display.
\end{remark}

\subsection{Certification around a computed solution}\label{sec:certificate}
Theorem~\ref{thm:local} guarantees convergence from starts
close enough to a minimum-edge representation, but the size of that
neighborhood depends on unknown quantities, so the theorem cannot tell
whether a particular computed output is reliable. We therefore give a
check that is run after the algorithm stops. It takes the computed point
$\bar x$ and a small ball around it and verifies two properties on the
ball. First, one cycle of the algorithm keeps the support of $\bar x$.
Second, with this support fixed, one cycle is a contraction. If both
hold, the contraction mapping theorem gives a fixed point of the
algorithm inside the ball, every start in the ball converges to it, and
the computed point lies within a known distance of it. Checks of this
kind are standard in validated numerics, where interval arithmetic turns
floating-point computations into rigorous bounds, for instance in
interval Newton methods for nonlinear equations
\citep{moore2009,rump}. The difference here is that one cycle involves
discrete choices, namely the support and the vertex order, and the
conditions (C1)--(C6) below control them.

Let $\bar x=(\bar B,\bar\Omega)$ be a feasible computed point, with candidate directed-edge support $E$, and let $N$ index its candidate bidirected support together with all diagonal entries of $\Omega$. Write $\Pi_E,\Pi_N$
for the associated coordinate projections, and define the following map
using the observed covariance $S$:
\begin{align*}
 Z_B(x)&=B-\alpha\nabla_B\ell_S(B,\Omega),&U(x)&=\Pi_E Z_B(x),\\
 Z_\Omega(x)&=\Omega-\beta\nabla_\Omega\ell_S(U(x),\Omega),&
 \mathcal T(x)&=(U(x),\Pi_N Z_\Omega(x)).
\end{align*}
Under the conditions below, $\mathcal T$ is the smooth formula for a
complete cycle; it extends to cyclic coefficient inputs
with positive definite error covariance, which permits derivative bounds
on a convex neighborhood.

Use orthonormal coordinates for the product Frobenius norm, write
$\|\cdot\|_2$ for the Euclidean norm of a vector and the spectral norm of
a matrix, choose an
invertible matrix $Q$, and consider the ellipsoid
$\mathcal B=\{x:\|Q(x-\bar x)\|_2\le r\}$. Require positive definite input error
covariances throughout $\mathcal B$, and impose the following strict
conditions uniformly over this ellipsoid:
\begin{enumerate}
\item[(C1)] Active directed-edge inputs satisfy
$\delta<|Z_{B,ij}|<M_B$ for every $(i,j)\in E$.
\item[(C2)] Inactive directed edges that can individually be added to
$E$ without creating a directed cycle satisfy
$|Z_{B,ij}|<2\alpha\lambda/\delta$.
\item[(C3)] The inputs corresponding to active bidirected edges have
magnitude exceeding $\delta$.
\item[(C4)] Inactive off-diagonal inputs to the $\Omega$-step have
magnitude below $\beta\lambda/\delta$.
\item[(C5)] The projected error covariance satisfies
$mI\prec\Pi_N Z_\Omega\prec M_OI$.
\item[(C6)] Every order that does not extend $E$ has strictly
smaller total scalar proximal gain than the orders that extend $E$, so the full
$B$-step chooses the candidate directed-edge support.
\end{enumerate}
A compatible extra directed edge is removed by its scalar threshold
in (C2); an inactive edge that would create a directed cycle with $E$
cannot be removed in this way, and (C6) excludes it through the comparison
of orders. We check (C1)--(C6) and the bounds of the next theorem by
interval arithmetic, with a two-state subset dynamic program for (C6)
(Section~S17 of the Supplementary Material).

\begin{theorem}[A posteriori convergence certificate]\label{thm:posterior}
Fix $S\succ0$, a feasible center $\bar x$, positive tuning parameters, and
$0<\alpha<\delta^2/(2\lambda)$ and $0<\beta<\delta^2/\lambda$ for the stated proximal updates.
Suppose conditions (C1)--(C6) hold on $\mathcal B$, and
verified bounds give
\[
 \sup_{x\in\mathcal B}\|QD\mathcal T(x)Q^{-1}\|_2\le q<1,\qquad
 \|Q(\mathcal T(\bar x)-\bar x)\|_2\le e_0,\qquad e_0+qr\le r.
\]
Then the following conclusions hold.
\begin{enumerate}
\item[(i)] The exact full alternating map has a unique fixed point
$x^\dagger$ in $\mathcal B$, and this point is a limiting critical point
of the full objective.
\item[(ii)] Every feasible initialization in the ball remains there and
identifies the candidate support after one cycle.
\item[(iii)] The iterates converge geometrically, and the distance from
the center to the fixed point is bounded as follows:
\[
 \|Q(x^t-x^\dagger)\|_2\le q^t\|Q(x^0-x^\dagger)\|_2,
 \qquad \|Q(\bar x-x^\dagger)\|_2\le e_0/(1-q).
\]
\end{enumerate}
\end{theorem}

The proof is the contraction argument described at the start
of this subsection; conditions (C1)--(C6) ensure that one cycle equals
the smooth map $\mathcal T$ on the ellipsoid (Section~S17 of the
Supplementary Material).

\begin{remark}[Scope of the certificate]\label{rem:certificate_scope}
Theorem~\ref{thm:posterior} certifies local convergence of the
algorithm to the fixed point $x^\dagger$ and bounds the distance from
$\bar x$ to $x^\dagger$. Global optimality and membership in the
generating class rest on the statistical results of
Sections~\ref{sec:estimation} and~\ref{sec:separation}.
Since the ellipsoid includes nonzero inactive coordinates, the
conclusion covers nearby feasible starts with additional edges.
\end{remark}

Near a regular solution of Theorem~\ref{thm:local} with strict
support, order and eigenvalue margins, the conditions of
Theorem~\ref{thm:posterior} hold on a small ball in a suitably weighted
metric. Consequently, every sufficiently accurate computed center with the same
support passes the certificate once the interval bounds are fine enough
(Section~S17 of the Supplementary Material).
\section{Numerical experiments}\label{sec:numerics}
Our statistical results concern global minimizers, and our
guarantees for the algorithm are local. We check both on small graphs,
where the global minimizer can be found by fitting every support. In
Section~\ref{sec:global} we compare the selected support with the
minimum-edge fiber on three vertices. In Section~\ref{sec:verma} we check,
on four vertices and at the population level, that the minimum-edge
fiber respects an equality constraint that is not a conditional
independence. In Section~\ref{sec:random_starts} we run the alternating
algorithm from random starts on the three-vertex designs, and in
Section~\ref{sec:local_numerics} we illustrate its local convergence and
the refit rate. Section~S18 of the Supplementary Material
\citep{supplement} gives the numerical details and a check of the
certificate of Theorem~\ref{thm:posterior} on a two-variable example. Python and R
code that reproduces every computation in this section is available at
\url{https://github.com/AsiaeeLab/sparsest-representations-hidden-confounding}.

\subsection{Global selection on three vertices}\label{sec:global}
Theorems~\ref{thm:ident} and~\ref{thm:shrinking} predict that, with
$\lambda_n=\delta_n=\log(n)/n$, the support of the thresholded global
minimizer of the saturated criterion is, with probability tending to
one, a minimum-edge support, whose graph lies in the generating class.
On three vertices we can fit every acyclic mixed support, so we can
compare the selected support with the minimum-edge fiber sample by
sample and see how quickly this limit is reached. There are 200 acyclic mixed supports, obtained by combining each of the
25 acyclic directed graphs with each of the 8 sets of bidirected
edges, bows included. For every support $H$ and sample we compute the
constrained likelihood minimum $\ell_{\min}(H;S_n)$ over $\K$ with the
zeros of $H$ imposed, and we select the support that minimizes
$\ell_{\min}(H;S_n)+\lambda_ne(H)$.

This selection agrees with the global minimizer of the
saturated criterion when two inequalities hold. Because $q_\delta\le1$, the minimum of $\ell_{S_n}+\lambda_nP_{\delta_n}$
never exceeds the selected score, and the refit on the selected support
attains that score, since on every sample all its edge coordinates exceeded
$5\delta_n$ in magnitude. On every sample we checked the reverse
inequality, that no parameter with small unsaturated coordinates scores
lower, by local descent of the criterion from the six best refits, which
never found a smaller value. On a subset of 86 samples, consisting of
the 16 samples of the chain and the confounded pair whose selected
support is not a minimum-edge support, 20 front-door samples that
underfit and 50 others, we also certified the computation with outward-rounded interval arithmetic and branch and
bound. Across these 86 samples, all but two of the 17{,}200 support fits are within $10^{-8}$ of the exact constrained minima, and the remaining two cannot change the selection. On 84 of these samples, the minimum of the criterion of Theorem~\ref{thm:shrinking} over $\K$ is within $2\times10^{-8}$ of the selected score, so the selected refit is a global minimizer to that accuracy (Section~S18 of the Supplementary Material).
On the remaining samples the support fits use multistart local
optimization, and the selected refit is a numerically supported
candidate for the global minimizer. Thresholding at
$\delta_n/2$ never changed the selected support.

We use three generating graphs with unit error variances and the
domain $|B_{ij}|\le2$, $0.05I\preceq\Omega\preceq10I$. They are the chain
$1\to2\to3$ of Example~\ref{ex:three_vertex}(a) with $B_{12}=0.8$ and
$B_{23}=-0.7$; the confounded pair $1\to2\leftrightarrow3$ with
$B_{12}=0.7$ and $\Omega_{23}=0.5$; and the front-door graph with
$1\to2\to3$ and $1\leftrightarrow3$, $B_{12}=0.7$, $B_{23}=0.6$ and
$\Omega_{13}=0.4$. Table~\ref{tab:census} describes their
minimum-edge fibers. The first two have $e^*=2$, and their minimum-edge
fibers lie on five and four supports, two and three of them with
bidirected edges. The front-door graph is complete and bow-free, so its
covariance model contains every positive definite matrix. Its minimum-edge fiber lies on all 37 three-edge supports of
full dimension, 12 of them bowed, and selection there tests the edge count
alone. For each $n\in\{100,200,400,\dots,6400\}$ we draw 200 samples.

\begin{figure}[t]
\centering
\includegraphics[width=\linewidth]{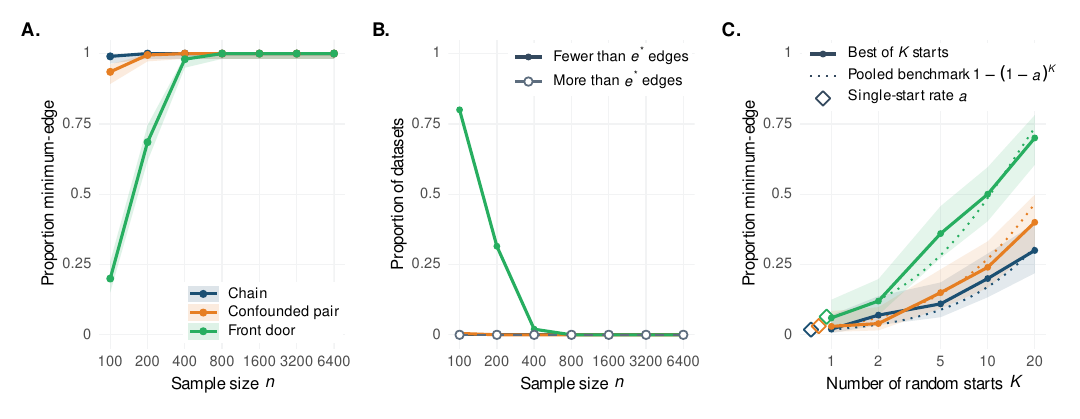}
\caption{\textbf{Global Selection and Random Starts on Three Vertices.}
\textbf{A.} Proportion of the 200 samples whose selected support,
thresholded at $\delta_n/2$, is a minimum-edge support, for
$\lambda_n=\delta_n=\log(n)/n$, with Wilson 95\% bands.
\textbf{B.} Proportion of samples selecting fewer than $e^*$ edges (solid
lines) or more than $e^*$ edges (gray dashed line, zero for every design
and $n$); the remaining errors have $e^*$ edges in another algebraic class.
\textbf{C.} At $n=1000$, proportion of the 100 samples in which the
lowest-objective run among the first $K$ random starts of
Algorithm~\ref{alg:estimator} ends at a minimum-edge support, for
$\lambda=0.01$ and $\delta=0.1$, with Wilson 95\% bands. Diamonds to the
left of $K=1$ mark the single-start rate $a$ estimated from 2{,}000 starts
pooled over the samples, and dotted curves show the pooled-rate
benchmark $1-(1-a)^K$.}
\label{fig:global}
\end{figure}

For the chain and the confounded pair, the selected support is a
minimum-edge support in 198 and 187 of the 200 samples at $n=100$ and in
every sample from $n=400$ on, while for the front-door graph it is one
in 40 samples at $n=100$ and in every sample from $n=800$ on
(Table~\ref{tab:census} and
Figure~\ref{fig:global}\textbf{A}--\textbf{B}). No sample selected more
than $e^*$ edges, and since on these designs the minimum-edge fibers lie on
exactly the $e^*$-edge supports of the generating classes, the same
proportions are also the rates of class selection. Fifteen samples of the first two designs, 13 of them for the
confounded pair, which has the smaller gap, selected
$e^*$ edges in another algebraic class. The other erroneous selections had
too few edges, and all but one were in the front-door design, which underfits at
small $n$. Its smallest likelihood gap in Table~\ref{tab:census} is $0.037$, attained when dropping one edge. Since $\log(n)/n$ falls below $0.037$ only for $n\ge132$, at $n=100$ even the population criterion prefers two edges, and $\lambda_n$ exceeds every admissible $\lambda_0$ of Theorem~\ref{thm:penalty}.

\begin{table}[t]
\centering
\caption{\textbf{Minimum-Edge Fibers and Global Selection on Three Vertices.}
For each design, the number of supports in the minimum-edge fiber and
how many of them carry a bidirected edge or a bow; the smallest
likelihood gap $\ell_{\min}(H;\Sigma^*)-\ell_*$ over supports $H$ with at
most $e^*$ edges that do not represent $\Sigma^*$ exactly; and the
proportion of the 200 samples whose selected support, thresholded at
$\delta_n/2$, is a minimum-edge support, for
$\lambda_n=\delta_n=\log(n)/n$. Every proportion equals one for
$n\ge800$.}
\label{tab:census}
\begin{tabular}{lcccccccc}
\toprule
 & & \multicolumn{3}{c}{Minimum-edge supports} & & \multicolumn{3}{c}{Selection at $n=$}\\
\cmidrule(lr){3-5}\cmidrule(lr){7-9}
Design & $e^*$ & All & Bidirected & Bowed & Smallest gap & 100 & 200 & 400\\
\midrule
Chain & 2 & 5 & 2 & 0 & 0.096 & 0.990 & 1.000 & 1.000\\
Confounded pair & 2 & 4 & 3 & 0 & 0.052 & 0.935 & 0.995 & 1.000\\
Front-door & 3 & 37 & 31 & 12 & 0.037 & 0.200 & 0.685 & 0.980\\
\bottomrule
\end{tabular}
\end{table}

\subsection{A constraint beyond conditional independence}\label{sec:verma}
The designs of Section~\ref{sec:global} test identification only through
conditional independences. On four vertices, an algebraic class can be
strictly smaller than a Markov class. The bow-free graph $G^*$ with edges
$1\to2\to3\to4$ and $2\leftrightarrow4$ carries the Verma constraint, an
equality that is not a conditional independence \citep{nested2023}. Its
covariance model satisfies $X_1\perp X_3\mid X_2$ and the cubic equality
\[
 \sigma_{13}\sigma_{22}\sigma_{34}-\sigma_{14}\sigma_{22}\sigma_{33}
 -\sigma_{13}\sigma_{23}\sigma_{24}+\sigma_{14}\sigma_{23}^2=0,
\]
and its variety has dimension $p+e^*=8$ (Lemma~\ref{lem:dimension}). The
graph with edges $3\to2\to1\to4$ and $2\leftrightarrow4$ also has four
edges and the single conditional independence $X_1\perp X_3\mid X_2$, so it
is Markov equivalent to $G^*$, but it satisfies the cubic with vertices 1
and 3 exchanged and lies in another algebraic class.

At a randomly drawn parameter of $G^*$, we fitted every one of the
34{,}752 acyclic mixed supports on four vertices to the population
covariance. The minimum exact edge count is $e^*=4$, and exactly two
supports with four edges represent $\Sigma^*$, namely $G^*$ and the graph with
edges $1\leftrightarrow2$, $2\to3$, $3\to4$ and $2\leftrightarrow4$. Both
lie in the algebraic class of $G^*$, as Theorem~\ref{thm:ident}
predicts. The Markov-equivalent graph above, and its variant with
$3\leftrightarrow2$ in place of $3\to2$, do not represent $\Sigma^*$, and
the best fits found for them have likelihood gap about $0.725$. For each
of the other supports with at most four edges, nonexactness does not
rest on the fits, because a vanishing partial correlation or a relabeled cubic
that holds throughout the model of the support but fails at $\Sigma^*$
excludes exact representations with any parameter values (Section~S18 of the
Supplementary Material). Joint sparsity therefore selects the algebraic
class of $G^*$ and excludes graphs that share its conditional
independences.

\subsection{Random starts}\label{sec:random_starts}
From an arbitrary start we can guarantee only a critical
point (Theorem~\ref{thm:optimization}), and a minimum-edge representation
only from a nearby start (Theorem~\ref{thm:local}); the saturated
criterion moreover has local minima at bowed representations
(Section~\ref{sec:l1}). We ran Algorithm~\ref{alg:estimator} on the three designs of
Section~\ref{sec:global}, with 100 samples for each
$n\in\{250,1000,4000\}$ and fixed tuning parameters $\lambda=0.01$ and
$\delta=0.1$, for which a multistart computation indicates that the
population minimizers are exactly the minimum-edge representations.
Each sample received 20 random starts, with coefficients and
error-correlation perturbations of scale $0.5$, and each run was stopped
after at most 20{,}000 cycles (Section~S18 of the Supplementary Material
gives the step sizes). Continuing the random runs that reached this
budget until they met the stopping tolerance turned one failure into a
success and no success into a failure, so we report the outcomes within
the budget.

A single start ended at a minimum-edge support in only about $2\%$, $3$--$4\%$ and $6$--$7\%$ of runs for the chain, the confounded pair and
the front-door graph, with no trend in $n$. Every other run ended with
more than $e^*$ edges, almost always at a bowed support; the
converged failures examined in Section~S18 of the Supplementary Material
were non-strict local minima of the criterion, at which the same mechanism as in
Section~\ref{sec:l1} traps a local algorithm. The best of
the 20 random starts by objective value ended at a minimum-edge support in $29$--$30\%$, $40$--$53\%$ and $70$--$80\%$ of
samples, close to the pooled-rate benchmark $1-(1-a)^{20}$ for the single-start rate $a$
(Figure~\ref{fig:global}\textbf{C}). Because this benchmark ignores differences in difficulty across samples, it tends to overstate success. Starts ten times smaller, with
coordinates mostly below $\delta$, raised the best-of-20 proportions to $97$--$100\%$, $73$--$97\%$ and
$100\%$ (30 samples per cell). The start $B=0$,
$\Omega=\operatorname{diag}(S_n)$ succeeded in almost every sample for the
chain and the front-door graph and in none for the confounded pair, where
it usually ended at a complete directed acyclic graph. With the 20 scale-$0.5$ starts, this diagonal start, and starts at the refits of the 50 best-scoring supports that enumeration provides here, the lowest objective across all 71 runs was attained at a minimum-edge support in 897 of the 900 samples.

\subsection{Local convergence and refitting}\label{sec:local_numerics}
\begin{figure}[t]
\centering
\includegraphics[width=\linewidth]{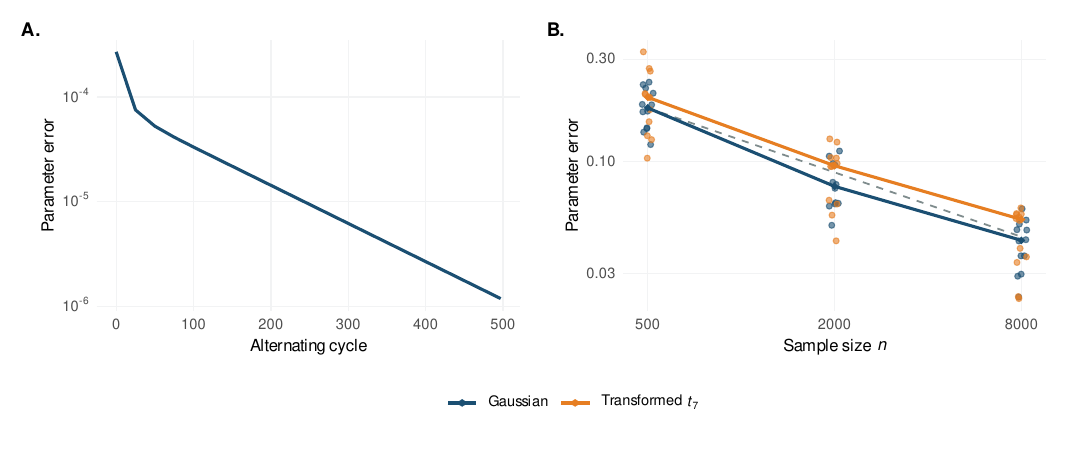}
\caption{\textbf{Alternating Convergence and Refitting Near a Fixed Representation.}
\textbf{A.} Population parameter error along the alternating sequence for
the four-variable model of Section~\ref{sec:local_numerics},
illustrating contraction near the fixed representation.
\textbf{B.} Individual sample refit errors and their medians for 12
Gaussian and 12 transformed-$t_7$ samples at each of $n=500,2000,8000$, with an
$n^{-1/2}$ guide anchored at the Gaussian median for $n=500$.
Each run starts from a perturbation of the population parameters and uses
a local likelihood refit.}
\label{fig:alternating}
\end{figure}

To illustrate Theorem~\ref{thm:local} and the refit rate of
Theorem~\ref{thm:transfer}(iii), we start the algorithm near a fixed
representation of a four-variable model with $B_{12}=0.9$, $B_{13}=0.8$,
$B_{23}=-0.75$ and $B_{34}=0.85$, unit error variances, and
$\Omega_{14}=0.3$.
We perturb the population parameters with standard deviation
$10^{-4}$ and use $\lambda=0.008$, $\delta=0.15$, $\alpha=0.02$,
$\beta=0.08$, $|B_{ij}|\le2$ and eigenvalue constraints $0.1I\preceq\Omega\preceq4I$. For each of $n=500,2000,8000$ we draw 12
Gaussian samples and 12 linear transforms of standardized $t_7$ samples
with the same covariance, run the alternating algorithm, threshold at
$\delta$ and refit the likelihood locally on the selected support.
All 72 final supports, with and without thresholding, match the
fixed representation, and the support is identified after one cycle in
66 runs, with all six exceptions at $n=500$. The median refit errors decrease at
roughly the $n^{-1/2}$ rate, from 0.178 to 0.0429 for Gaussian samples and
from 0.200 to 0.0538 for transformed-$t_7$ samples
(Figure~\ref{fig:alternating}).

\section{Discussion}\label{sec:discussion}
When errors are correlated, the covariance does not determine the
representation, and we have used the joint edge count of directed and
bidirected edges to choose among the representations. Generically, the
sparsest representations form a finite target whose graphs lie in the
generating algebraic class, a saturated penalty selects this target in
the population and from data, and an $\ell_1$ penalty can miss even the
Markov class. The alternating algorithm is guaranteed only locally, so
its output should be compared across starts and checked with the
a posteriori certificate.

The statistical results hold for fixed $p$ and concern global
minimizers, with constants depending on $\Sigma^*$, and their generic
identification conclusion requires a generating graph whose
covariance variety has dimension $p+e^*$ (Remark~\ref{rem:bows}).
The fixed-parameter thresholds $\delta_0$ and $\lambda_0$ of Theorem~\ref{thm:penalty} are nonconstructive and depend on $\Sigma^*$, and our exact $B$-step takes time exponential in $p$
(Section~\ref{sec:steps}). The alternating procedure has
no guarantee of reaching a minimum-edge representation from an arbitrary
start. Our numerical experiments use three and four
variables, where the global minimizer can be computed; the behavior of
the algorithm on larger graphs remains to be studied.

\textit{Using the selected class.} The graphs in the selected
class impose the same conditional independences and other polynomial
equalities on the covariance (Section~\ref{sec:verma}) but can differ in edge type and orientation
(Example~\ref{ex:three_vertex}). A causal
conclusion from one selected graph is therefore supported by the data only
when it holds for every graph of the class that can represent the
covariance. When the selected
member is bow-free, the algorithms of \citet{ommen2024} test its
algebraic equivalence with other bow-free graphs, and the implied causal
effects can be compared, in the spirit of \citet{nowzohour}. 
\textit{Practical recommendations.} For continuous-optimization learners of mixed graphs, our theoretical and numerical results support a saturated penalty, thresholding at a fixed fraction of $\delta_n$ followed by refitting, and a BIC-order schedule such as $\lambda_n=\delta_n=\log(n)/n$; the random-start study indicates that several starting points are needed, and the certificate of Theorem~\ref{thm:posterior} can then be applied to the selected limit.

\textit{Calibration of the tuning parameters.} The window of tuning parameters for
class selection in Section~\ref{sec:separation_setting} is explicit, but
its endpoints involve the separation $\Delta$ and the minimum edge count
$e$, which depend on $\Sigma^*$. A calibration of
$\delta$ and $\lambda$ from observable quantities that retains the
selection conclusions would make this window
usable in practice.

\textit{Dependence on dimension.} By
Proposition~\ref{prop:fano}, no estimator selects a correlated pair
uniformly unless $nr^2$ is at least of order $\log p$, and the sufficient
condition of Theorem~\ref{thm:certificate} involves
$L_\Phi$ and $b^2/a$, which grow with $p$. Closing this gap requires a sharper description of the alternatives
that control selection.

\textit{Scalable coefficient steps and initialization.} Replacing the
exact order search by a scalable $B$-step that retains the objective
decrease and the relative residual bounds of
Proposition~\ref{prop:inexact} would remove the exponential cost.
Conditions under which a practical initializer enters the neighborhood of
Theorem~\ref{thm:local} are equally important; an $\ell_1$ fit
need not enter it (Section~\ref{sec:l1}).

\textit{Pervasive confounding.} A different modeling choice treats
hidden confounding as pervasive and low rank. \citet{cpw2012} estimate a
sparse conditional graph and a low-rank latent component by convex
optimization, and \citet{poet} estimate a covariance matrix as such a
component plus a sparse remainder. \citet{frot} and \citet{rsvp} remove
such a component before structure learning, and \citet{decam} do so for
nonlinear models, building on the representation of margins of Bayesian
networks by \citet{evans}. In identification theory, \citet{barber2022} directly treat latent factors that load on many variables and note that
bidirected edges are natural when each hidden cause affects only a few
observed variables, the regime of this paper. In multivariate regression,
coefficients and a residual precision matrix are estimated jointly
\citep{mrce,lin}, and \citet{hive} remove hidden-variable effects from the
coefficients through the low-rank part of the residual covariance; there
predictors and responses are separate sets, while in \eqref{eq:model}
every coordinate is a response. Combining a few pervasive factors with
sparse error correlations is a natural extension, pursued via
precision-matrix decomposition in \citet{asiaee2025}.

\textit{Extending the certificate.} A version of the
certificate of Theorem~\ref{thm:posterior} that verifies entry from a
larger set of starting points, checking the support
changes and order comparisons of the full algorithm, would connect
the local guarantee to the global question.

\begin{funding}
AA and SP were supported in part by the Patient-Centered Outcomes
Research Institute (PCORI) under award ME-2023C1-32148. AA was also
supported in part by the National Human Genome Research Institute under
award R00HG011367 and by the National Institute of Mental Health under award
R01MH139379, both of the National Institutes of Health.
\end{funding}

\begin{supplement}
\stitle{Supplement to ``Identifying and estimating sparsest representations of linear Gaussian causal models with hidden confounding''}
\sdescription{Proofs of all numbered results, the $\ell_1$
calculations, the certification procedure and the numerical details
(Section~S18).}
\end{supplement}
\begin{supplement}
\stitle{Code}
\sdescription{Python and R code that reproduces the numerical results,
tables and figures of Section~\ref{sec:numerics} and of Sections S13 and S18 of the
Supplementary Material, also available at
\url{https://github.com/AsiaeeLab/sparsest-representations-hidden-confounding}.}
\end{supplement}

\clearpage
\includepdf[pages=-]{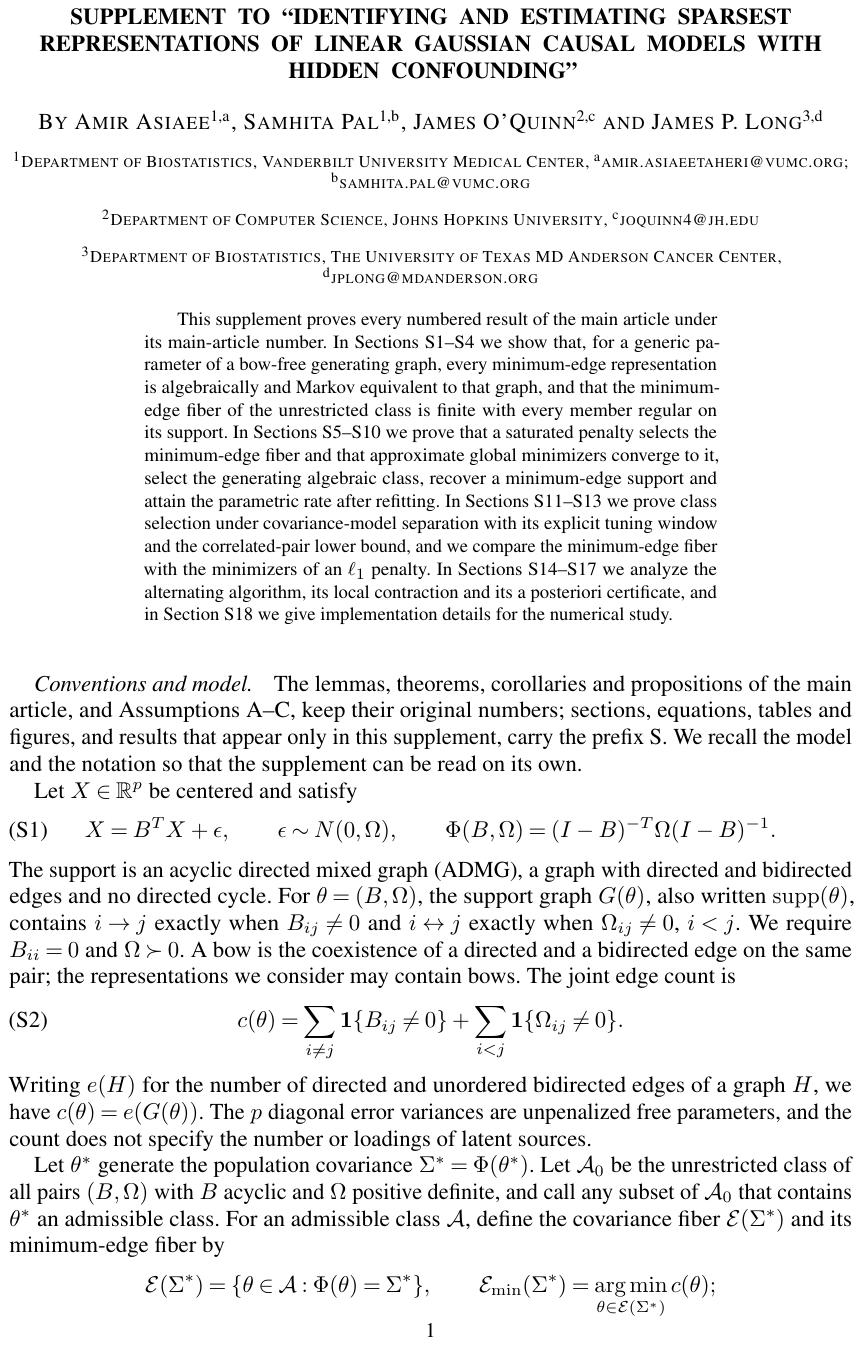}
\end{document}